\documentclass[a4paper,USenglish,cleveref, thm-restate]{lipics-v2021}

\usepackage[T1]{fontenc}
\usepackage[utf8]{inputenc}
\usepackage{lmodern}
\usepackage{amsmath,amssymb,amsthm}
\usepackage{microtype}
\usepackage{booktabs,array}
\usepackage{enumitem}
\usepackage{xcolor}
\usepackage{tikz}
\usetikzlibrary{arrows.meta,positioning,calc,decorations.pathreplacing,fit,patterns}
\usepackage{caption}
\usepackage{cleveref}

\newcommand{\bits}{\{0,1\}}                 
\newcommand{\g}[1]{\mathsf{#1}}             

\newcommand{\qstart}{q_{\mathrm{start}}}

\newcommand{\sem}[1]{[\![#1]\!]}            
\newcommand{\upd}[2]{[#1{\leftarrow}#2]}    

\newcommand{\ins}[1]{\textsf{#1}}           
\newcommand{\nop}{\mathsf{nop}}
\newcommand{\push}{\mathsf{push}}
\newcommand{\pop}{\mathsf{pop}}

\definecolor{pathc}{HTML}{CFE3D4}
\definecolor{posc}{HTML}{DDE6F2}
\definecolor{pc}{HTML}{DDE6F2}
\definecolor{qc}{HTML}{CFE3D4}
\definecolor{dc}{HTML}{F6DCCB}
\definecolor{edit}{HTML}{B5532A}
\definecolor{ink}{HTML}{2F4B7C}
\definecolor{accent}{HTML}{B5532A}
\definecolor{greyl}{HTML}{888888}

\title{Pushdown Reachability Requires $N^{2}$ Time on Multitape Turing Machines}
\author{A. R.~Balasubramanian}{Max Planck Institute for Software Systems}{bayikudi@mpi-sws.org}{}{}
\author{Rupak Majumdar}{Max Planck Institute for Software Systems}{rupak@mpi-sws.org}{}{}
\authorrunning{A. R. Balasubramanian and R. Majumdar}



\nolinenumbers
\begin{document}
\maketitle

\begin{abstract}
Pushdown reachability, or equivalently, language non-emptiness of pushdown automata (PDA), is a fundamental problem in automata theory and program analysis. We show that, for every $\epsilon >0$, no deterministic multitape Turing machine can decide pushdown reachability in $O(N^{2-\epsilon})$ time, where $N$ is the input size. 
The lower bound already holds for \emph{sparse instances}, i.e., instances in which the number of transitions
is linear in the number of states. This gives the first unconditional super-linear lower bound for pushdown reachability. Furthermore, we also extend this result and show that for any $\epsilon > 0$ and $k \ge 1$, no deterministic multitape Turing machine can decide the intersection non-emptiness problem of one PDA and $k-1$ NFAs in time $O(N^{2k-\epsilon})$, where $N$ is the input size.

Finally, pushdown reachability is well-known to be quasi-linear time equivalent to many different problems in program analysis. Our result translates to all of those problems as well and provides the first unconditional super-linear lower bounds for those problems on multitape Turing machines.
\end{abstract}

\section{Introduction}\label{sec:intro}
Proving super-linear lower bounds for natural computational problems
remains a central challenge in complexity theory. 
In this paper, we prove such a lower bound for pushdown reachability: 
for every $\epsilon>0$, we prove that no deterministic multitape Turing machine can decide the
problem in $O(N^{2-\epsilon})$ time, where $N$ is the input size. 

Pushdown reachability is a standard and well-studied problem in the analysis of recursive programs. 
A pushdown automaton (PDA) consists of a finite control graph and a stack; each transition
changes the control state and pushes a symbol, pops a symbol, or
leaves the stack unchanged. 
Given two control states $s,t$, the reachability
problem asks whether some run starting at $s$ with an empty stack reaches $t$.  
Reachability with such constraints underlies precise interprocedural dataflow
analysis~\cite{RHS95} and symbolic verification of pushdown
systems~\cite{BEM97,ABEGRY05}; in fact, the problem is quasi-linear-time equivalent (i.e., equivalent
up to reductions running in time $O(n \log^c n)$ for some $c$) to a large number of problems from program analysis \cite{HeintzeMcAllester,Neal,CMS22}. 

For any fixed stack alphabet, the standard saturation algorithms solve pushdown reachability in cubic
time in the number of control states. It is known that this cubic upper bound can be improved by polylogarithmic factors~\cite{Rytter85,Chaudhuri08,Majumdar26}. However, despite decades of research on this problem, no substantially better algorithms are known, i.e., no algorithm running in time $O(n^{3-\epsilon})$ for any $\epsilon > 0$ is known. 

%

\paragraph*{Lower bounds.}
Time hierarchy theorems, beginning with Hartmanis and
Stearns~\cite{HartmanisStearns65}, establish that more computational time
allows us to solve strictly more problems. Such \emph{unconditional lower bounds} are obtained by diagonalization.
However, transferring these bounds to a particular algorithmic problem remains difficult and there are very few
unconditional lower bounds known for problems in $\mathsf P$, or indeed for problems in $\mathsf{NP}$ or even $\mathsf{PSPACE}$. A survey from 1995 \cite{Regan95} captures the state of the art then, but not much has changed since.
%

Paul, Pippenger, Szemer\'edi, and Trotter~\cite{PPST83}
separated deterministic from nondeterministic linear time on multitape
Turing machines. 
Their proof shows an $\Omega(n(\log^* n)^{\frac14})$ lower bound on any deterministic Turing machine
accepting any language complete for $\mathsf{NTIME}(n)$ under deterministic linear-time many-one reductions.
Grandjean \cite{Grandjean88,Grandjean90} showed $\mathsf{NTIME}(n)$-hardness for 
some natural $\mathsf{NP}$-complete problems
from automata theory, thus showing an unconditional $\Omega(n(\log^* n)^{\frac14})$ lower bound for 
these problems.

For problems in $\mathsf P$, the landscape is bleak.
Adachi, Iwata, and Kasai~\cite{AIK84} obtained $\Omega(n^k)$ lower bounds 
for families of combinatorial pebble games that belong to
$\mathsf P$ when the number of pebbles is fixed.
Cook \cite{Cook71} showed that auxiliary pushdown automata with $c k \log n$ auxiliary store simulate $\mathsf{DTIME}(n^k)$ for some constant $c$,
so that their emptiness problems inherit unconditional lower bounds.
Emptiness of auxiliary-store pushdown machines can be simulated by the intersection non-emptiness problem
for one pushdown automaton and $k-1$ deterministic finite automata (DFAs).
Swernofsky and Wehar~\cite{SW15} proved unconditional 
lower bounds for this problem. 
Their result states that there exists a constant $c$ (that covers the cost of reductions from deterministic
Turing machines to the intersection non-emptiness problem) such that 
intersection non-emptiness of a pushdown automaton and $k-1$ DFAs cannot be solved in deterministic
time $n^{ck}$.
The exact value of $c$ is not specified; in particular, their proof does not show a 
non-trivial lower bound for pushdown reachability.

Recent progress on space-efficient simulation gives another route to
unconditional time lower bounds. 
Williams~\cite{Williams25} showed that
 multitape Turing machines running in time $T$ can be simulated using
$O(\sqrt{T\log T})$ space. 
Combining this simulation with a space-hierarchy
argument, Wehar~\cite[Theorem~8]{Wehar26} rules out deterministic multitape
algorithms taking
$
  o\!\left(\frac{L^2}{(\log L)^3(\log\log L)^2}\right)
$
time for the intersection non-emptiness problem of an input list of DFAs, where $L$ is
the total binary encoding length. Since the number of DFAs in that problem is part of
the input, the problem is $\mathsf{PSPACE}$-complete.  
This result does not subsume our bounds for a single pushdown system or for one PDA and
each fixed number of DFAs, which are polynomial-time problems. 
%


\paragraph*{Our results.}
We prove an unconditional polynomial lower bound for pushdown reachability
on deterministic multitape Turing machines.

\begin{theorem}[Lower Bound for Pushdown Reachability]\label{intro:thm:reach}
For every $\epsilon>0$, no deterministic multitape Turing machine
can decide pushdown reachability in time $O(N^{2-\epsilon})$,
where $N$ is the size of the input. Moreover, the lower bound already holds for \emph{sparse instances}, i.e.,
when the number of transitions is linear in the number of states.
\end{theorem}

To the best of our knowledge, this provides the first unconditional super-linear lower bound for pushdown reachability, 
even though the problem has been studied since the 1960s.  Furthermore, this lower bound also translates to many problems from program analysis such as data-flow reachability, control-flow reachability, GMR-reachability as well as 2NPDA recognition and CFL reachability~\cite{HeintzeMcAllester,CMS22}, since all of these problems are quasi-linear-time equivalent to pushdown reachability.\footnote{Technically speaking, in these papers, this quasi-linear time equivalence has been shown over the word RAM model, and not over the multitape Turing machine model. However, by looking at the reductions, one can easily see that this equivalence also holds for the multitape Turing machine model.}


Note that pushdown reachability is easily seen to be equivalent to the language non-emptiness problem for pushdown automata. Viewed this way, we generalize this problem and we also consider the intersection non-emptiness for one PDA
and any fixed number of NFAs. More precisely, in this problem, we are given a PDA $A_1$ and a fixed number (say $k-1$) of NFAs $A_2,\ldots,A_{k}$. We then have to decide whether there exists a word in
$\mathcal L(A_1)\cap\bigcap_{i=2}^{k}\mathcal L(A_i)$. 
The standard way to solve this problem is to first do a product construction of the NFAs with the PDA
and then solve pushdown reachability on the resulting PDA. For constant input alphabets, 
this gives an $O(N^{3k})$ algorithm. Balasubramanian, Chistikov, and Majumdar~\cite{BCM25} studied the fine-grained complexity of this problem, giving lower bounds based on the (since refuted) clique hypothesis.
We obtain the following unconditional exponent.

\begin{theorem}[Lower Bounds for Intersection Non-Emptiness of PDA and $(k-1)$ NFAs]\label{intro:thm:intersection}
For every fixed $k\ge1$ and every $\epsilon>0$, no deterministic
multitape Turing machine decides intersection non-emptiness of one PDA and
$k-1$ NFAs in time $O\!\left(N^{2k-\epsilon}\right)$.
\end{theorem}

In particular, for the case of one PDA and one NFA, the theorem shows that this problem cannot be solved in time $O(N^{4-\epsilon})$ for any $\epsilon > 0$. Compared with the $n^{\Omega(k)}$ lower bounds of Swernofsky and Wehar~\cite{SW15}, our
construction identifies the exponent $2k$ and also yields a
super-linear bound for the base problem, i.e., pushdown reachability. 

\paragraph*{Proof Idea: From Turing machines to small PDAs.}
On a very high level, our proof gives a way to succinctly represent computations of a Turing machine running in time $T$
by a PDA of size roughly $T^{1/2+o(1)}$. To this end, our proof consists of three reductions.

In the first reduction, we revisit the classical simulation of Turing machines running in time $T$ by circuits of size $O(T \log T)$~\cite{PF79}. We examine this construction and observe that this circuit is so-called \emph{locally-addressable} and can be described very succinctly, i.e., exponentially more succinctly than $O(T \log T)$. To this end, we show that each gate of the circuit can be named by a type (of $O(\log \log T)$ bits) from a ``small'' set of types and an address of
\[
  H=\lceil\log_2(8T)\rceil
\]
bits. The type and the address together uniquely determine a gate. Furthermore, we also show that given any gate,
its predecessors in the circuit can be computed by small \emph{Boolean programs}, i.e., finite-state machines which can edit a fixed number of bits. Such a circuit is obtained essentially by a recursive simulation of the given Turing machine, following the construction discussed by Shalunov~\cite{Sha25}. At each recursive level $j$, we simulate $2^j$ different computations, each taking $O(T/2^j)$ time. A naive way to reference each cell of each computation of each level $j$ is to use $2H$ bits: the first $H$ bits to store which computation in level $j$ we are referencing, and the next $H$ bits to store which cell in that computation we are referencing. However, we note that the number of cells that we need to consider in a level halves every time we go down a level. Using this observation, we show that each cell in each level $j$ can be captured by $H$ bits: $j$ bits to capture which computation and the remaining $H-j$ bits to capture the position of the cell in that computation.
It is extremely important that we are able to do this using $H = \log T + O(1)$ bits and not using $O(\log T)$ bits.
As we shall see in the proof, encoding each cell in each level using the latter would not give us the required lower bound! 

In the second reduction, we translate these locally-addressable circuits into an intermediate language called
\emph{relation programs}. Roughly speaking, relation programs comprise a finite set of variables, each of which
denotes a set of $H$-bit strings. Each variable $A$ is determined by a unique equation such as $A := [u_t = c]$ (the set of $A$ is the set of strings whose $t^{th}$ bit is $c$) or $A := A_1 \lor A_2$ (the set of $A$ is the union of the sets of $A_1$ and $A_2$) or $A := A_1 \land A_2$ (the set of $A$ is the intersection of the sets of $A_1$ and $A_2$). Importantly, a variable $A$ can occur in the right-hand side of its own equation such as $A := A \lor B$. In this case,
the semantics is taken to be the least fixed point that is arrived at by starting with the empty set for each variable.
We show how to translate a locally-addressable circuit of size $S$, $H$ address bits and $m$ inputs into a relation program of size $O(S+mH)$. 

In the third reduction, we compile the relation program into a PDA. To this end, we divide an $H$-bit string into 2 equal-sized blocks of strings of size $H/2$ each. For each variable $A$, we construct a \emph{procedure} $P_A$ (i.e., a gadget PDA) whose states are of the form $(\ell,r)$ where $\ell$ is a location (from a ``small'' set of locations)
and $r$ is a bit string of length $H/2 + O(\sqrt H)$, which is used to store an encoding $\varphi(x)$ of a bit string $x$ of length $H/2$. This procedure $P_A$ will have the property that a string $xy$ will belong to the set of $A$ iff $P_A$ can reach the state $(\ell_{end},\varphi(y))$ from the state $(\ell_{start},\varphi(x))$ (here $\ell_{start}$ and $\ell_{end}$ will be two specific locations of $P_A$ called its entry and exit). Hence, while storing only one of the two blocks of an $H$-bit string at any time, $P_A$ can still be used to verify whether a given $H$-bit string
belongs to the set of a given variable. The main difficulty is the case $A := A_1 \land A_2$: here the procedures of $A_1$ and $A_2$ have to be started from the same $x$ and have to end with the same $y$, although the state has no room for both $x$ and $y$. We solve this by choosing the encoding $\varphi$ so that it contains no three-term arithmetic progression, adapting a construction of Behrend~\cite{Behrend46}; this allows the PDA to keep $\varphi(x)$ and $\varphi(y)$ in its state \emph{as a sum} and still check, with the help of the stack, that the procedures of $A_1$ and $A_2$ start and end on the same blocks. We show that combining all of the $P_A$ together gives us a PDA of size 
\[
  O\!\left(|\Pi| \log|\Pi| \cdot H \log H \cdot 2^{H/2+O(\sqrt H)}\right),
\]
where $|\Pi|$ is the size of the relation program. 

Combining these three reductions together (with a careful counting of the computation time) would then show that if pushdown reachability can be solved in time $O(N^{2-\epsilon})$, then the original Turing machine can be simulated in time less than $T$. Intuitively, this is because the exponential factor in the PDA in the third reduction is $2^{H/2+O(\sqrt H)}$ and 
$H = \log T + O(1)$ (and not $c \log T$ for some $c > 1$). Since we can show that this applies to any arbitrary Turing machine running in time $T$, this would contradict the time hierarchy theorem. We can therefore conclude the desired lower bound for pushdown reachability.

For the intersection non-emptiness problem of one PDA and $(k-1)$ NFAs, we do the same three reductions as before to get a PDA in the third reduction that stores (an encoding of) one of the two blocks of a string of $H$ bits. We then show that we can further split this block into $k$ smaller blocks, encode each of them separately, and distribute one such encoded block each to a smaller PDA and $(k-1)$ many smaller NFAs.
This reduces the exponential factor in each of these machines to $2^{H/2k+O(\sqrt H)}$. We then apply a similar argument to conclude the desired lower bound for the intersection non-emptiness problem.

\paragraph*{Conditional bounds and fine-grained complexity.}
Much recent progress on the running times of problems in
$\mathsf P$ has come from fine-grained reductions. Such reductions relate
the possibility of faster algorithms for many problems to hypotheses about
a few source problems. 
For example, under the NFA acceptance hypothesis \cite{NFA}, there is no truly
subcubic algorithm for pushdown reachability.
Unfortunately, several conditional bounds in fine-grained complexity used hypotheses
such as the exponential time hypothesis or the clique hypothesis which have now 
been refuted \cite{ETHOpenAI,ETHAnthropic,AW26,WilliamsZ26}.
Thus, lower bounds for pushdown reachability based on these hypotheses are void.


\section{Preliminaries}\label{sec:prelims}
In this section, we recall the formal definitions of NFAs and PDAs and state our main result, namely an $N^{2k}$ lower bound for the intersection non-emptiness problem of one PDA and $(k-1)$ many NFAs. Throughout this section, let $\Sigma$ be a finite \emph{input alphabet} and $\Gamma$ be a finite \emph{stack alphabet}.

\begin{definition}[NFA]\label{pre:def:nfa}
A \emph{nondeterministic finite automaton} (NFA) is a tuple $(Q,\Sigma,\delta,q_0,F)$ where $Q$ is a finite set of \emph{states}, $q_0 \in Q$ is the \emph{initial state}, $F \subseteq Q$ is a set of \emph{final states}, 
and $\delta \subseteq Q \times \Sigma \times Q$ is a finite set of \emph{transitions}.
\end{definition}

We will often denote a transition $(q,a,q')$ as $q\xrightarrow{a}q'$. A run of an NFA on a word $w = a_1a_2\cdots a_n \in \Sigma^*$ is a sequence of transitions of the form $q_0 \xrightarrow{a_1} q_1 \xrightarrow{a_2} q_2 \dots \xrightarrow{a_n} q_n$. This run is said to be accepting if $q_n \in F$ and in that case we say that $w$ is accepted by the NFA. The language of an NFA is the set of all words that it accepts. 

\begin{definition}[PDA]\label{pre:def:pda}
A \emph{pushdown automaton} (PDA) is a tuple $(Q,\Sigma,\Gamma,\delta,q_0,q_f)$ where $Q$, $q_0$ and $q_f$ are the set of states, the initial state and the final state respectively, and $\delta \subseteq Q \times \Sigma \times \{\nop, \push\,\gamma, \pop\,\gamma : \gamma \in \Gamma\} \times Q$ is a finite set of transitions.
\end{definition}

As before, we will denote a transition $(q,a,\mathit{op},q')$ as $q \xrightarrow{a \mid \mathit{op}} q'$, where
we will say that $\mathit{op}$ is the \emph{stack operation} of the transition. Intuitively, such a transition allows the PDA to read $a$ and move from $q$ to $q'$ while performing the operation $\mathit{op}$ on its stack. 

Formally, a \emph{configuration} of a PDA is a pair $(s,w)$ where $s$ is a state and  $w\in\Gamma^*$ is the stack content; the top of the stack is the last letter of $w$. A transition $q\xrightarrow{a\,\mid\,\mathit{op}}q'$ allows a \emph{step} from $(q,w)$ to $(q',w')$ that reads $a$ if
\begin{itemize}
\item $\mathit{op}=\nop$ and $w'=w$ (the stack is not changed), or
\item $\mathit{op}=\push\,\gamma$ and $w'=w\gamma$ (the letter $\gamma$ is put on top), or
\item $\mathit{op}=\pop\,\gamma$ and $w=w'\gamma$ (the top letter is $\gamma$, and it is removed).
\end{itemize}
The initial configuration of a PDA is the configuration $(q_0,\epsilon)$. Similar to an NFA, we can define runs of a PDA over its configurations. We say that a run is accepting if it ends at the configuration $(q_f,\epsilon)$. The \emph{language} of a PDA is the set of words that it accepts.

The \emph{size} of an NFA or a PDA is the number of its states plus the number of its transitions. We say that a family of NFAs or PDAs is \emph{sparse} if the number of transitions of each machine in this family is a linear function of its number of states. 

We are now ready to state the central problem that we shall consider in this paper, namely the \emph{intersection non-emptiness problem}. To this end, fix a $k\ge1$. The intersection non-emptiness problem for one PDA and $k-1$ NFAs is the following problem:
\begin{quote}
\emph{Given:} A PDA $A_1$ and NFAs $A_2,\dots,A_k$, all over the same input alphabet $\Sigma$.\\
\emph{Decide:} Is there a word that is accepted by all of $A_1,\dots,A_k$?
\end{quote}

For $k=1$, there are no NFAs, and the problem asks whether the language of a given PDA is non-empty. The input letters play no role for this question. So this is the \emph{pushdown reachability problem}: can a given PDA go from its initial state with the empty stack to the final state with the empty stack?

The central result of this paper is the following.

\begin{theorem}[Main Theorem]\label{fin:thm:main}
For any $k \ge 1$ and any $\epsilon>0$, there is no deterministic multitape Turing machine that solves the intersection non-emptiness problem for one PDA and $k-1$ NFAs in time $O(N^{2k-\epsilon})$, where $N$ is the maximum size of the given machines. Moreover, this lower bound holds even if all the machines are sparse and the input alphabet and stack alphabet are of constant size.
\end{theorem}

In particular, for $k=1$ we obtain the following corollary.

\begin{corollary}\label{fin:cor:reach}
For any $\epsilon>0$, there is no deterministic multitape Turing machine that solves the pushdown reachability problem in time $O(N^{2-\epsilon})$, where $N$ is the size of the given PDA. Moreover, this lower bound holds even when the PDAs are sparse and the stack alphabet is of constant size.
\end{corollary}

\begin{remark}
    Note here that we have stated our results in terms of the maximum size of the machines in the given input, where the size is taken to be the number of states and transitions of the given machines. Technically speaking, when a machine is encoded as input, it needs $O(N \log N)$ bits where $N$ is its size. Hence, ideally, we would want our result to rule out $O((N \log N)^{2k-\epsilon})$ time algorithms for every $\epsilon > 0$, because $O(N \log N)$ is the size of the input. However, notice that ruling out $O(N^{2k-\epsilon})$ time algorithms for every $\epsilon > 0$ automatically rules out $O((N \log N)^{2k-\epsilon})$ time algorithms for every $\epsilon > 0$ as well. 
    Hence, for the rest of the paper, we will concentrate on proving the theorems as they are stated here, with the understanding that the same result applies when we instead want our lower bound to be a function of the input size.
    Similarly, since the lower bound holds even for sparse machines, it also follows that the same lower bound applies
    as a function of the number of states as well.
\end{remark}

The rest of the paper is structured as follows. The proof of \cref{fin:thm:main} goes through three reductions. In \cref{sec:tm} we show that a computation of a Turing machine of length $T$ can be simulated by a Boolean circuit whose gates have names of $\log T+O(1)$ bits, such that the names of the predecessors of a gate can be computed from its name by very simple programs; we call such circuits \emph{locally addressable}. In \cref{sec:cr} we translate locally addressable circuits into \emph{relation programs}, a simple formalism of equations between sets of strings over $\{0,1\}$. In \cref{sec:int} we then translate relation programs into instances of the intersection non-emptiness problem. Finally, in \cref{sec:fin} we combine the three reductions with the time hierarchy theorem and prove \cref{fin:thm:main}.

\section{From Turing Machines to Locally Addressable Circuits}\label{sec:tm}
In this section we shall see how to simulate a Turing machine running in time $T$ by a Boolean circuit of size $O(T \log T)$. This is a classical result of Pippenger and Fischer~\cite{PF79}.
However, here we revisit that result and define a similar (but technically different) construction which additionally satisfies the property of being \emph{locally addressable}. Roughly speaking, this means that the circuit is extremely structured: Each gate $g$ has a unique name representable using $\log T + O(\log\log T)$ bits. Furthermore, the names of each of the predecessors of $g$ can be computed from its name by a simple \emph{Boolean program}. Such a program is just a finite-state automaton that can additionally edit the bits of the name of the gate $g$. Thus the names of the predecessors of $g$ are obtained \emph{by editing its name in place} with no other additional memory. 

For the rest of this section, we assume that the reader is familiar with the basic notions of Turing machines and Boolean circuits. In particular, in our circuits we allow input gates, constant gates (i.e., gates whose output is either the constant 0 or the constant 1) and the usual $\lor, \land$ and $\lnot$ gates of fan-in at most 2. We begin by formalizing the notion of a locally addressable circuit.


\subsection{Locally Addressable Circuits}\label{tm:sec:goal}

In order to define locally addressable circuits, we need the notion of a Boolean program.

\begin{definition}[Boolean programs]\label{tm:def:aprog}
A \emph{Boolean program} (on $H$ Boolean variables) is a tuple $(Q,\delta,q_0,F)$ where $Q, q_0, F$ are a set of states, the initial state and a set of final states respectively, and $\delta$ is an instruction set that, for every state $p \notin F$, contains exactly one instruction of the form $\textsf{test}(i;p_0,p_1)$ or $\textsf{set}(i,c;p')$, where $1 \le i \le H$, $p_0,p_1,p' \in Q$ and $c \in \{0,1\}$.
\end{definition}

Intuitively, a Boolean program contains $H$ Boolean variables $1,\dots,H$. At each point, a configuration of such a machine is given by a state $q$ and a binary string $x_1\cdots x_{H}$ with the interpretation that the $i^{th}$ bit $x_i$ is stored in the $i^{th}$ Boolean variable. From such a configuration, an instruction of the form $\textsf{test}(i;p_0,p_1)$ allows the Boolean program to move to state $p_0$ if $x_i = 0$ and to $p_1$ otherwise. Similarly,
an instruction of the form $\textsf{set}(i,c;p')$ allows the program to update the variable $x_i$ to $c$ and move to $p'$. Final states do not have any instructions.

The initial configuration is given by the initial state $q_{0}$ and some input string $z \in \{0,1\}^H$. Starting from this configuration, if the computation of the Boolean program reaches a configuration of the form $(q_f,x)$ with $q_f \in F$, we say that this initial configuration is accepted and that its output is $(q_f,x)$.

Having defined Boolean programs, we can now define locally addressable circuits.



\begin{definition}[Locally addressable circuits]\label{tm:def:lac}
Let $C$ be a Boolean circuit with $m$ inputs.
We say that $C$ is a locally addressable circuit with a local description of \emph{address length} $H$ if the following conditions are satisfied:
\begin{enumerate}[label=(\alph*)]
\item \emph{The name condition.} Each gate of $C$ is uniquely identified by a pair 
$(\tau,z)$ where $\tau$ comes from a finite set $\mathcal T$ of \emph{types} and 
$z \in \{0,1\}^H$.  $\mathcal T$ contains two special types $\mathsf{in}, \mathsf{out}$ such that the $i^{th}$ input gate of $C$ is $(\mathsf{in},i)$ and the output gate of $C$ is $(\mathsf{out},0)$. Additionally, any two gates of the same type perform the same operation (i.e., input, constant 0, constant 1, $\land, \lor, \lnot$).

\item \emph{The predecessor condition.} Each type $\tau \in \mathcal T$ has two Boolean programs $f_{\tau,0}$ and $f_{\tau,1}$ such that 1) the initial state of both these programs is $\tau$, 2) the final states of both these programs belong to $\mathcal T$ and 3) for a gate $(\tau,z)$, its $i^{th}$ predecessor is given by the output of $f_{\tau,i}$ on $(\tau,z)$.

\end{enumerate}

The \emph{size}  $S$ of the local description is the number of types and the number of states and instructions in each Boolean program $f_{\tau,i}$. 
\end{definition}

The intuitive idea is that any gate $g$ of $C$ can be uniquely identified by a tuple $(\tau,z)$ where $\tau$ comes from a ``small'' set of types and $z$ is some binary string of length $H$. The predecessors of $g$ can be computed from $(\tau,z)$ by a Boolean program that depends only on the type $\tau$. Of course, every $C$ has a local description for a large enough $H$ or a large enough $S$. The main point of our construction is that from a Turing machine running in time $T$, we can construct a locally addressable circuit of address length $H = \log T + O(1)$ and 
size $S = \text{poly}(H) = \text{polylog}(T)$. We stress that $H$ will be  $\log T + O(1)$
and not $O(\log T)$ and this will be extremely crucial to us later on.
More formally, we prove the following.

\begin{theorem}\label{tm:thm:main}
Let $M$ be a deterministic multitape Turing machine running in time $T(m) \ge m$. 
Let $m \in \mathbb{N}, T = T(m)$ and $H=\lceil\log(8T)\rceil$. Given $m, T$ in binary,
we can compute, in $\text{poly}(H)$ time, a local description of a circuit $C$ with address length $H$ and size $\text{poly}(H)$ such that for any input $x \in \{0,1\}^m$, $M$ accepts $x$ iff the output of $C$ on $x$ is 1.
\end{theorem}

As mentioned before, the important point is the address length $H$ which is of the form $\log T+O(1)$. As we shall see later on, this will be extremely important for the later reductions, as even an address length of $2 \log T$ will not work for the lower bound. Furthermore, since the local description of $C$ has size $\text{poly}(H) = \text{polylog}(T)$, its number of types is $\text{polylog}(T)$ and so its total number of gates will be $2^{H} \cdot \text{polylog}(T) = T \text{polylog}(T)$. (Actually it will turn out that the polylog$(T)$ term is just $O(\log T)$).

The rest of this section is devoted to proving Theorem~\ref{tm:thm:main}. This proof is split into three parts.
In the first part, we recall the known recursive simulation of a $T(m)$-time Turing machine using $O(T \log T)$ steps. 
In the second part, we show how to convert this recursive simulation into a Boolean circuit. Finally, in the third part, we show that this circuit is locally addressable with address length $H=\lceil\log(8T)\rceil$.

\subsection{Part 1: Recursive Simulation}\label{tm:sec:window}

In this section, we recall the known recursive simulation of the computation of a Turing machine on some input word $x \in \{0,1\}^m$ for $T$ many time steps, roughly as described by Shalunov~\cite[Appendix~B]{Sha25}, who attributes it to Kouck\'y. For our purposes, we assume that the given Turing machine $M$ has some $k$ (bidirectionally) infinite tapes, with the first tape being the input tape. On input $x\in\bits^m$, the machine
starts in its initial state $\qstart$ with its heads in cell 0, the bits $x_1\cdots x_{m}$ in cells
$0,\dots,m-1$ of the input tape and blanks elsewhere. A configuration of such a machine is a tuple containing the current state of the machine, and for each tape, the cell in which the head of that tape is located and the contents of the tape. The semantics of such a machine is taken to be the usual well-known semantics on its configurations given by means of its transition function.




A configuration of size $n$ is a configuration in which each tape is restricted to just $n$ cells (which will be denoted by the positions $-n/2,-n/2+1,\dots,n/2-1$). Any tape of such a configuration can be split into 8 consecutive blocks of equal size, which we call \emph{tape blocks}, numbered from 0 to 7. We say that a configuration is centered (resp. semi-centered) if the head of each tape belongs to the $i^{th}$ block for $i \in \{2,3,4,5\}$ (resp. $i \in \{1,2,3,4,5,6\}$).

\begin{figure}[t]
\centering
\begin{tikzpicture}[font=\scriptsize,>={Stealth[length=1.8mm]}]
\foreach \y/\s/\h in {2.4/4/5.5,1.2/1/2.3,0/4/6.6} {
  \foreach \i in {0,...,7} {\draw (\i*1.1,\y) rectangle ++(1.1,0.5); \node at (\i*1.1+0.55,\y+0.25) {\i};}
  \fill[posc,opacity=0.7] (\s*1.1,\y) rectangle ++(4.4,0.5);
  \foreach \i in {0,...,7} {\node at (\i*1.1+0.55,\y+0.25) {\i};}
  \draw[very thick] (\s*1.1,\y) rectangle ++(4.4,0.5);
  \draw[<-,thick,edit] (\h*1.1,\y+0.5)--(\h*1.1,\y+0.82);
}
\node[anchor=west] at (9.1,2.65) {head in block $5$: window $4$--$7$};
\node[anchor=west] at (9.1,1.45) {head in block $2$: window $1$--$4$};
\node[anchor=west] at (9.1,0.25) {head in block $6$: window $4$--$7$};
\draw[decorate,decoration={brace,mirror,amplitude=3pt}] (1.1,-0.1)--(2.2,-0.1) node[midway,below=3pt] {$n/8$};
\end{tikzpicture}
\caption{Choosing the window (shaded) from the block of the head (arrow). The window never extends
beyond the array, and the head lies in its central half.}
\label{tm:fig:window}
\end{figure}
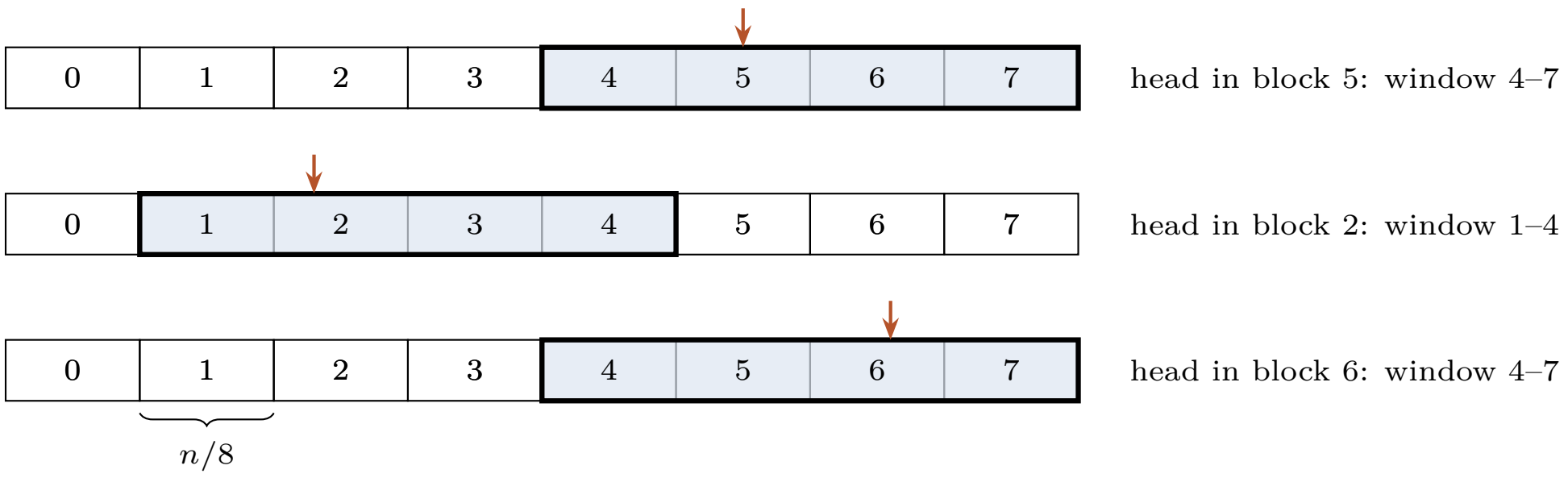

From a semi-centered configuration $U$ of size $n$, we can extract a centered configuration $V$ of size $n/2$, by just considering the adjacent blocks of each head of each tape. More precisely, if the head of the $i^{th}$ tape belongs to block $s(i)$, then we construct $V$ by restricting $U$ in each of its tapes $i$ to the blocks $s'(i)-1,s'(i),s'(i)+1,s'(i)+2$ where $s'(i) = \min(5,s(i))$.
(The cell numbers of $V$ in tape $i$ will now be renumbered by subtracting $(s'(i)-3) n/8$ from its original number in $U$ to lie in the range $-n/4,\dots,n/4-1$; see \cref{tm:fig:window}.)

With these definitions in hand, the recursive simulation of $M$ can now be stated as follows:
Suppose we want to simulate $M$ from a centered configuration $U$ of size $8n$ for $n$ steps.
If $n = 1$, we directly do the simulation. Otherwise, from $U$, we first extract the centered configuration $V_1$ of size $4n$ and recursively simulate the machine from $V_1$ for $n/2$ steps. (Note that in $n/2$ steps, the heads will never move beyond $n/2$ cells in either direction and so the heads will never leave these $4n$ cells in $V_1$). Let $W_1$ be the configuration that is reached at the end of this simulation. Then, in $U$, we replace the contents of the cells corresponding to $V_1$ with the corresponding cells of $W_1$. We also replace the state and the head positions of $U$ by the ones given by $W_1$. Let $X$ be this new semi-centered configuration. 

From $X$, we again extract the centered configuration $V_2$ of size $4n$ and recursively simulate the machine from $V_2$ for $n/2$ steps. Let $W_2$ be the configuration that is reached. Once again, in $X$, we replace the contents of the cells corresponding to $V_2$ and also the control state and the heads with the ones given by $W_2$ to get a new configuration $Y$. It is then easily verified that $Y$ is the configuration that is reached from $U$ after $n$ steps. This completes the description of the recursive simulation.

Note that this recursion can be naturally seen as a tree where the root simulates the machine on the initial configuration for $T$ steps (we can restrict the initial configuration to only take cells in the range $-4T,\dots,4T-1$ with the input written on cells $0,\dots,m-1$ of the input tape). Furthermore, each node at level $j$ simulates $M$ on a centered configuration of size $8T/2^j$ for $T/2^j$ steps. Note that a node at level $j$ of the tree can be uniquely labelled by a string from $\{0,1\}^j$ obtained by following the path from the root to that node.

\subsection{Part 2: The Circuit}\label{tm:sec:circuit}

We now implement the recursion described above by a circuit. To this end, we fix some reasonable encoding of a configuration of size $n$ as a string of length $O(n)$: the state and the content of each cell can be encoded using a fixed number (say $r$) of bits. Furthermore, we assume that the encoding also records, for each cell, whether the head is currently present at that cell or not, by a special \emph{head bit} (which we take to be the last bit).

Fix a node $u \in \{0,1\}^j$ of the recursion tree at some level $j$. Recall that it represents the simulation of the machine on a configuration of size $8n := 8T/2^j$ for $n := T/2^j$ steps. More precisely, $u$ can be thought of as receiving a configuration $X$ of size $8n$ as input and outputting 
the configuration $Y$ of size $8n$ that will be reached by $X$ in $n$ steps. 
With this intuition, for each $\ell \in \{-4n,\dots,4n-1\}$, we will associate a constant number of gates that are responsible for receiving the content of the $\ell^{th}$ cell of some tape of $X$ as input and outputting the content of the $\ell^{th}$ cell of that tape of $Y$ as the output. (We will also have gates for inputting and outputting the state).

We now describe the gates associated to the node $u$ and the cell $\ell$ of a tape $e$.

\paragraph*{Sub-Part I: The input.} The gates $\g{In}^{e,1}[u,\ell],\dots,\g{In}^{e,r}[u,\ell]$ carry the $r$ bits of the content of cell $\ell$ of tape $e$ in $X$. In addition, the gates $\g{InState}^{1}[u],\dots,\g{InState}^{r}[u]$ carry the state of $X$. 

\paragraph*{Sub-Part II: Finding the head.}
For each block $s\in\{1,\dots,6\}$, a gate $\g{Head}^{e}_s[u]$ computes the disjunction of the head bits $\g{In}^{e,r}[u,\ell']$ over all cells $\ell'$ of block $s$. Hence $\g{Head}^{e}_s[u]$ is $1$ if and only if the head of tape $e$ is in block $s$. (Technically, we allow circuits to only have fan-in 2, whereas such a gate requires unbounded fan-in. But this can be easily rectified by the usual way of replacing a gate with arbitrary fan-in by a collection of bounded fan-in gates as follows:
$
\g{Tree}^{e}_0[u,\ell]=\g{In}^{e,r}[u,\ell]$,$
\g{Tree}^{e}_h[u,\ell]=\g{Tree}^{e}_{h-1}[u,\ell]\vee\g{Tree}^{e}_{h-1}[u,\ell+2^{h-1}]$,
with $\g{Head}^e_s[u]$ finally set to $\g{Tree}^{e}_{\log n}[u,\ell_s]$, where $\ell_s=(s-4)n$ is the first cell of block $s$. This does not change any of the constructions or results below, but just adds more notation).

Recall that if the head is in block $s$, we have to take the blocks $s'-1,\dots,s'+2$ with $s'=\min(5,s)$ and send them to the next recursive step. Furthermore, while sending them, we have to renumber each cell in these blocks
by subtracting $c\cdot n$, where $c:=s'-3\in\{-2,-1,0,1,2\}$. Correspondingly, we introduce one gate for each possible value of $c$:
\begin{gather*}
\g{Sel}^{e}_{-2}[u]=\g{Head}^{e}_1[u],\qquad
\g{Sel}^{e}_{-1}[u]=\g{Head}^{e}_2[u],\qquad
\g{Sel}^{e}_{0}[u]=\g{Head}^{e}_3[u],\\
\g{Sel}^{e}_{1}[u]=\g{Head}^{e}_4[u],\qquad
\g{Sel}^{e}_{2}[u]=\g{Head}^{e}_5[u]\vee\g{Head}^{e}_6[u].
\end{gather*}
If $X$ is semi-centered, exactly one of these five \emph{selectors} is $1$, and it tells us the value of $c$. 

\paragraph*{Sub-Part III: Calling the recursion.} 
We now have to send the selected four blocks along with the shifted cell numbers to the next recursive step, i.e., to the node $u0$ as input. Correspondingly, for every cell $-2n\le\ell<2n$,
every tape $e$, and every $1 \le i \le r$, we set
\[
\g{Cand}^{e,i}_c[u0,\ell]=\g{Sel}^{e}_c[u]\wedge\g{In}^{e,i}[u,\ell+c n]\quad(c=-2,\dots,2),
\qquad
\g{In}^{e,i}[u0,\ell]=\bigvee_{c=-2}^{2}\g{Cand}^{e,i}_c[u0,\ell].
\]

Note that since exactly one of the $\g{Sel}^e_c[u]$ is 1, $\g{In}^{e,i}[u0,\ell]$ is effectively $\g{In}^{e,i}[u,\ell+c n]$ for the $c$ which corresponds to the block where the head was present. Finally, along with the cell contents, the state is simply passed on:
$\g{InState}^{i}[u0]=\g{InState}^{i}[u]$.

Recursively, we now perform the same construction for the node $u0$ and construct output gates
$\g{Out}^{e,i}[u0,\ell]$ and $\g{OutState}^{i}[u0]$.

\paragraph*{Sub-Part IV: Putting the result back.} We now have to replace the blocks that we sent to $u0$ by the output of $u0$. Correspondingly, for every cell $-4n\le\ell<4n$, every tape $e$ and every $1 \le i \le r$, we set
\begin{align*}
\g{Ret}^{e,i}_c[u,\ell]&=
\begin{cases}
\g{Sel}^{e}_c[u]\wedge\g{Out}^{e,i}[u0,\ell-c n] & \text{if }-2n\le\ell-cn<2n,\\[2pt]
\g{Sel}^{e}_c[u]\wedge\g{In}^{e,i}[u,\ell] & \text{otherwise,}
\end{cases}
\qquad(c=-2,\dots,2),\\[4pt]
\g{Mid}^{e,i}[u,\ell]&=\bigvee_{c=-2}^{2}\g{Ret}^{e,i}_c[u,\ell].
\end{align*}

Similarly, we set $\g{MidState}^i[u]  = \g{OutState}^i[u0]$. By construction, the configuration encoded by the $\g{Mid}^{e,i}[u,\ell]$ and $\g{MidState}^i[u]$ gates is the configuration that is reached in $n/2$ steps from the configuration encoded by the $\g{In}^{e,i}[u,\ell]$ and
$\g{InState}^i[u]$ gates. 

\paragraph*{Sub-Part V: Repeating Sub-Parts I--IV once again.} Sub-Part V repeats Sub-Parts~I to~IV with $\g{Mid}[u]$ in place of $\g{In}[u]$, the $\g{MidState}[u]$ in place of $\g{InState}[u]$, and the child $u1$ in place of $u0$. The gates produced at the very end of this sub-part are the output gates $\g{Out}^{e,i}[u,\ell]$ and $\g{OutState}^{i}[u]$, representing the configuration that is reached in $n/2$ steps from the configuration represented by the $\g{Mid}[u]$ and $\g{MidState}[u]$ gates.
Combined with the previous sub-parts, this allows us to conclude that the output gates of $u$ represent the configuration that is reached in $n$ steps from the configuration represented by the input gates of $u$. It is immediate to see that the number of gates corresponding to $u$ is $O(n)$.

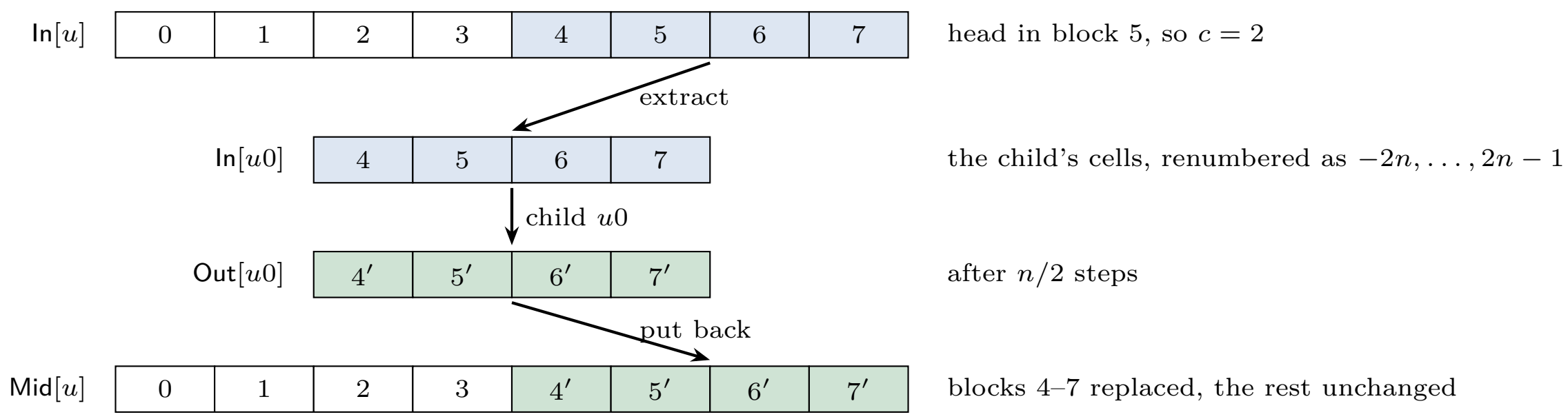
\begin{figure}[t]
\centering
\begin{tikzpicture}[font=\scriptsize,>={Stealth[length=1.8mm]}]
\def\w{0.95}
\node[anchor=east] at (-0.15,3.52) {$\g{In}[u]$};
\foreach \t/\c [count=\k from 0] in {0/white,1/white,2/white,3/white,4/posc,5/posc,6/posc,7/posc}
  {\draw[fill=\c] (\k*\w,3.3) rectangle ++(\w,0.44); \node at (\k*\w+\w/2,3.52) {\t};}
\node[anchor=west] at (7.85,3.52) {head in block $5$, so $c=2$};
\node[anchor=east] at (1.75,2.32) {$\g{In}[u0]$};
\foreach \t [count=\k from 2] in {4,5,6,7}
  {\draw[fill=posc] (\k*\w,2.1) rectangle ++(\w,0.44); \node at (\k*\w+\w/2,2.32) {\t};}
\node[anchor=west] at (7.85,2.32) {the child's cells, renumbered as $-2n,\dots,2n-1$};
\node[anchor=east] at (1.75,1.22) {$\g{Out}[u0]$};
\foreach \t [count=\k from 2] in {4$'$,5$'$,6$'$,7$'$}
  {\draw[fill=pathc] (\k*\w,1.0) rectangle ++(\w,0.44); \node at (\k*\w+\w/2,1.22) {\t};}
\node[anchor=west] at (7.85,1.22) {after $n/2$ steps};
\node[anchor=east] at (-0.15,0.12) {$\g{Mid}[u]$};
\foreach \t/\c [count=\k from 0] in {0/white,1/white,2/white,3/white,4$'$/pathc,5$'$/pathc,6$'$/pathc,7$'$/pathc}
  {\draw[fill=\c] (\k*\w,-0.1) rectangle ++(\w,0.44); \node at (\k*\w+\w/2,0.12) {\t};}
\node[anchor=west] at (7.85,0.12) {blocks $4$--$7$ replaced, the rest unchanged};
\draw[->,thick] (5.7,3.25)--(3.8,2.6) node[midway,right=4pt] {extract};
\draw[->,thick] (3.8,2.05)--(3.8,1.5) node[midway,right] {child $u0$};
\draw[->,thick] (3.8,0.95)--(5.7,0.4) node[midway,right=4pt] {put back};
\end{tikzpicture}
\caption{The first sub-parts of a node, drawn block by block. The numbers are the
block numbers of the node $u$; primed numbered blocks are the ones that were updated by the child $u0$.}
\label{tm:fig:stage}
\end{figure}

\paragraph*{Sub-Part VI: The base case.} Note that the base case of the leaves corresponds to the simulation of just a single step and can hence be represented by a constant number of gates for each node $u$ and each cell $\ell \in \{-4,\dots,3\}$. In fact, the gates for this circuit do not depend on $u$ at all and only depend on $\ell$ and the transition function of $M$. Hence, the circuits for all of the base cases are the same and are of constant size. This concludes the construction of the circuit.

It is then clear that the circuit constructed in this way has $O(T/2^j)$ gates for each node at level $j$ of the recursion tree and correctly computes the desired simulation. It remains to describe the input and output gates of the circuit. The circuit has $m$ input gates, where the $i^{th}$ input gate carries the bit $x_i$. The gates $\g{In}^{e,i}[u,\ell]$ and $\g{InState}^{i}[u]$ of the root node $u$ encode the initial configuration of $M$ on $x$. More precisely, for each cell $0 \le \ell < m$ of the input tape, the gates $\g{In}^{1,i}[u,\ell]$ compute the encoding of the symbol $x_{\ell+1}$ from the $(\ell+1)^{th}$ input gate; so each of them is either this input gate, or its negation, or a constant gate. All the other gates $\g{In}^{e,i}[u,\ell]$ and $\g{InState}^{i}[u]$ of the root node are constant gates: they encode the blank symbol in all other cells of all the tapes, the head of each tape at cell $0$, and the initial state. Finally, the circuit has a single output gate, which computes from the gates $\g{OutState}^{i}[u]$ of the root node whether the state reached after $T$ steps is the accepting state. Since there are $2^j$ nodes for each level $j$ and since the depth of the tree is $\log T$, the size of the circuit is $O(T \log T)$.

\subsection{Part 3: Local Addressability} 

We now show that the circuit described before admits a local description of address length $H = \lceil\log 8T\rceil$ and size $\text{poly}(H)$. To begin with, we start by defining the name of each gate of the circuit, i.e., the type and the address of each gate. First, note that the sub-circuit corresponding to a leaf node $u$ is of constant size and is the same for every leaf node $u$. Hence, we can let the type of any such gate be itself and its address be the string $u$ padded by the cell $\ell \in \{-4,\dots,3\}$ that the gate represents. $u$ requires $H-3$ bits to be encoded and $\ell$ requires 3 bits, giving the total to be $H$ bits. Note that we have only introduced a constant number of types so far.

Now, let $u \in \{0,1\}^j$ be a node at level $j$. Note that the computation that the node $u$ represents is over $8T/2^j = 2^{H-j}$ cells given by the range $\{-4T/2^j,\dots,4T/2^j-1\}$
and so each such cell $\ell$ can be represented using $H-j$ bits. Now, recall that every gate in the circuit corresponding to a node $u$ and a cell $\ell$ was either of the form
$\g{X}^{\,\cdot}_{\,\cdot}[u,\ell]$ or of the form $\g{X}^{\,\cdot}_{\,\cdot}[u]$. Hence, we can name the gates as follows.
\begin{itemize}
\item \emph{Address:} The address of $\g{X}^{\,\cdot}_{\,\cdot}[u,\ell]$ is taken to be
the $j$ bits of the node $u$ followed by the $H-j$ bits $\langle \ell \rangle$ of the
encoding of $\ell$ in two's-complement. The address of $\g{X}^{\,\cdot}_{\,\cdot}[u]$ is $u$ followed by $H-j$ zeros.
\item \emph{Type:} The type consists of the name $\g{X}$ ($\g{In}$, $\g{Sel}$, $\g{Cand}$,
\dots), the superscripts and subscripts (the tape $e$, the bit $i$, the shift $c$, the block $s$)
and the level $j$ of the node $u$. 
\end{itemize}

Note that every address has exactly $H$ bits. This also means that the same string can be the address of gates at different levels, but since the level is part of the type of a gate, this allows us to uniquely distinguish between different gates. Further note that the number of introduced types is just a polynomial in $H$. 

Having described the names of the gates, we now show how to compute the predecessors of the gates by Boolean programs. More precisely, for each type $\tau$, we have to devise Boolean programs $f_{\tau,0}$ and $f_{\tau,1}$ so that the $i^{th}$ predecessor of a gate $(\tau,z)$ can be obtained as the output of $f_{\tau,i}$. Going over the construction given in the previous subsection, this is immediately clear for the gates defined in Sub-Parts I, II, V and VI. In Sub-Part III, the main obstacle is to, for every level $j$ and each $c \in \{-2,\dots,2\}$, design a Boolean program to convert a bit string of the form $u 0 \langle \ell \rangle$ with $\ell \in \{-4T/2^{j+1},\dots,4T/2^{j+1}-1\}$ into a bit string of the form $u \langle \ell + c T/2^j\rangle$ where $c \in \{-2,\dots,2\}$. However, since $c T/2^j$ is a fixed number,
this is just adding a fixed number to an input number and this can be easily accomplished by a Boolean program with $O(H)$ states. Similarly, subtraction in Sub-Part IV can also be done.

Finally, we give the $i^{th}$ input gate the name $(\mathsf{in},i)$ and the output gate the name $(\mathsf{out},0)$. For the gates $\g{In}^{1,i}[u,\ell]$ of the root node, the Boolean program has to compare the cell $\ell$ with the fixed number $m$ and, if $0\le\ell<m$, replace $\ell$ by $\ell+1$ to reach the corresponding input gate; this can again be done with $O(H)$ states.
This completes the local description of the above circuit, (which can clearly be computed in time $\text{poly}(H)$), and thereby also the proof of Theorem~\ref{tm:thm:main}.

\section{From Locally Addressable Circuits to Relation Programs}\label{sec:cr}
In this section, we shall convert locally addressable circuits into an intermediate formalism that we call \emph{relation programs}. We shall first define relation programs formally and show how to compile a locally addressable circuit into a relation program. The main advantage of relation programs over locally addressable circuits is that we get a simple and clean formalism which is much easier to work with (for translation to pushdown automata). We now begin by introducing relation programs.

\subsection{Relation Programs}\label{cr:sec:relprog}

Throughout this section we consider strings $u$ of the form $u=u_1u_2\cdots u_{H} \in \{0,1\}^H$ for some $H$. For such a string $u$, a position $1 \le t \le H$ and a bit $c$, we write $u\upd{t}{c}$ for the string obtained from $u$ by setting bit $t$ to $c$.
 
\begin{definition}[Relation program]\label{cr:def:relprog}
A \emph{relation program} $\Pi$ over $H$ bits consists of a finite set $\mathcal V$ of \emph{variables}
and, for each variable $A\in\mathcal V$, exactly one \emph{equation} $A:=E$, where $E$ is of one of the
six forms
\[
\top,\qquad \bot,\qquad [u_t=c],\qquad A_1\vee A_2,\qquad A_1\wedge A_2,\qquad A_1\langle t\leftarrow c\rangle,
\]
with $A_1,A_2\in\mathcal V$, $1\le t\le H$ and $c\in\bits$. The \emph{size} of $\Pi$, denoted by $|\Pi|$, is its number of variables.
\end{definition}
 
Intuitively, every variable $A$ represents a subset $\sem A$ of $\{0,1\}^H$, and its equation $A := E$ says how this set $\sem A$ is defined:
\begin{itemize}
\item $A:=\top$: $\sem A$ is the entire set $\{0,1\}^H$.
\item $A:=\bot$: $\sem A$ is the empty set.
\item $A:=[u_t=c]$: $\sem A$ is the set of strings $u$ whose bit $u_t$ is $c$.
\item $A:=A_1\vee A_2$: $\sem A$ is the union of $\sem A_1$ and $\sem A_2$.
\item $A:=A_1\wedge A_2$: $\sem A$ is the intersection of $\sem A_1$ and $\sem A_2$.
\item $A:=A_1\langle t\leftarrow c\rangle$: $\sem A$ contains a string $u$ iff 
$u\upd{t}{c}$ is in $\sem A_1$.
\end{itemize}

While this is the intuitive idea, there is a small problem with this definition: An equation $A := E$ might have the variable $A$ in the right-hand side as well. In those cases, the above definition is circular. Therefore, we modify this definition so that we take the \emph{least fixed point} solution of this set of equations. Formally, we initially begin by setting $\sem A$ to be the empty set for all the variables and then apply the equations again and again until we reach a fixed point for all variables $A$. 

More precisely, let $X=(X_A)_{A\in\mathcal V}$ assign a subset $X_A \subseteq \{0,1\}^H$ to every variable $A$, and let $F_A(X)$ be the subset that the equation of $A$ produces from $X$, as described above; for instance $F_A(X)=X_{A_1}\cap X_{A_2}$ if the equation of $A$ is $A:=A_1\wedge A_2$, and $F_A(X)=\{u:u\upd tc\in X_{A_1}\}$ if the equation of $A$ is $A:=A_1\langle t\leftarrow c\rangle$. For $j\ge0$ we define the assignment $X^{j}$, the sets
\emph{after $j$ rounds}, by $X^{0}_A=\emptyset$ and $X^{j+1}_A=F_A(X^{j})$.

By definition of the equations, we can easily show that $X^{j}_A\subseteq X^{j+1}_A$
for all $j$. It follows then that after finitely many rounds, we reach a fixed point, i.e., there must be a round $j$ such that $X_A^{j} = X_A^{j+1}$ for all $A$. We then take $\sem A = X_A^{j}$ for that round $j$. 

For the sake of brevity, we will sometimes write compound expressions on the right-hand side of an equation, such as $A:=(A_1\wedge A_2)\vee A_3$ or $A:=A_1\vee A_2\vee A_3$. 
This is just an abbreviation: in such a case, every subexpression of such an equation gets an auxiliary variable of its own.

We can now state the main result of this section, which says that locally addressable circuits can be translated into small relation programs.

\begin{theorem}\label{cr:thm:main}
Let $C$ be a locally addressable circuit with $m$ inputs, address length $H$ and size $S$, and let $x \in \{0,1\}^m$ be an input. In $\text{poly}(S + mH)$ time, we can compute a relation program $\Pi$ over $H$ bits of size $O(S + mH)$ and a variable $R$ of $\Pi$ such that the output of $C$ on $x$ is $1$ if and only if $0^H \in \sem{R}$.
\end{theorem}

The rest of this section is devoted to proving this theorem.

\subsection{Proof of Theorem~\ref{cr:thm:main}}\label{cr:sec:translation}
 
Let $C$ be a locally addressable circuit with input $x=x_1\cdots x_{m}$, address length $H$ and size $S$. In particular, this means that every gate of $C$ can be represented
as $(\tau,u)$ where $\tau \in \mathcal{T}$ is a type and $u \in \{0,1\}^H$. 
The operation of the gate is entirely determined by its type, i.e., for any two $u, u'$,
the operations of $(\tau,u)$ and $(\tau,u')$ are the same. 
Furthermore, for $i \in \{0,1\}$, the $i^{th}$ predecessor of a gate $(\tau,u)$ in $C$
can be computed by the Boolean program $f_{\tau,i}$. 

For any gate $(\tau,u)$, we write $\mathrm{val}(\tau,u)$ for the value that it computes on the
input $x$. We now construct a relation program $\Pi$ over $H$ bits that simulates this circuit as follows.
 
\paragraph*{Part 1: The variables.} The program $\Pi$ will have the following variables.
\begin{itemize}
\item For every type $\tau$ and every $v\in\bits$, it will have a variable $A_{\tau,v}$. 
Intuitively, a string $u \in \{0,1\}^H$ will belong to $\sem{A_{\tau,v}}$ iff $\mathrm{val}(\tau,u) = v$.

\item For every type $\tau$, every $i, v \in\{0,1\}$ and every state $p$ of the Boolean program $f_{\tau,i}$, it will have a variable $B^{p,i}_{\tau,v}$. 
Intuitively, a string $u \in \{0,1\}^H$ will belong to $\sem{B^{p,i}_{\tau,v}}$ iff 
starting from the state $p$ with the Boolean variables set to $u$, $f_{\tau,i}$
accepts and its output is a gate whose value is $v$. In particular, if $p$ is the initial state,
then $u \in \sem{B_{\tau,v}^{p,i}}$ iff the $i^{th}$ predecessor of the gate $(\tau,u)$ has the value $v$.
\end{itemize}
 
\paragraph*{Part 2: The equations.} To begin with, we express the equations for the
variables $B_{\tau,v}^{p,i}$. We have three cases:
\begin{itemize}
\item Suppose $p$ is a final state. Recall that final states of the Boolean programs $f_{\tau,i}$ belong to the set of types $\mathcal{T}$. Hence $p \in \mathcal{T}$. 
In this case, according to the intuition above, we simply have to collect all
strings $u$ such that $(p,u)$ has value $v$ and this is precisely given by the variable 
$A_{p,v}$. Hence, we set 
\[
B_{\tau,v}^{p,i}:=A_{p,v}.
\]
\item Suppose $p$ has the instruction $\textsf{set}(t,c;p')$. Then from the state $p$, the Boolean program sets the $t^{th}$ bit to $c$ and moves to $p'$. Accordingly, we set
\[
B_{\tau,v}^{p,i}:=B_{\tau,v}^{p',i}\langle t\leftarrow c\rangle .
\]
\item Suppose $p$ has the instruction $\textsf{test}(t;p_0,p_1)$. Then from the state $p$, the Boolean program continues to either $p_0$ or $p_1$, depending on whether the $t^{th}$ bit is 0 or 1. Accordingly, we set
\[
B_{\tau,v}^{p,i}:=\big([u_t=0]\wedge B_{\tau,v}^{{p_0},i}\big)\vee\big([u_t=1]\wedge B_{\tau,v}^{{p_1},i}\big).
\]
\end{itemize}
 
Next we define the equations for the variables $A_{\tau,1}$ and $A_{\tau,0}$. Their equations entirely depend on the operation of the type $\tau$. First, for each Boolean program $f_{\tau,i}$ and every $v \in \{0,1\}$, we let $P_{\tau,v}^i$ denote the variable $B_{\tau,v}^{\tau,i}$ (recall that $\tau$ is the initial state of $f_{\tau,i}$). We can now define the equations as follows:
\begin{itemize}
\item Suppose the operation corresponding to $\tau$ is $\neg$. Then, the value of $(\tau,u)$
is $v$ iff the $0^{th}$ predecessor of $(\tau,u)$ has value $1-v$. Note that, by construction, $\sem{P_{\tau,(1-v)}^0}$ contains the string $u$ iff the $0^{th}$ predecessor of $(\tau,u)$ has value $1-v$. Hence, we can simply set $A_{\tau,v} := P_{\tau,(1-v)}^0$.

\item Suppose the operation corresponding to $\tau$ is $\wedge$. Then, the value of $(\tau,u)$
is $1$ iff both of its predecessors have value 1. Therefore, we can set
\[
A_{\tau,1}:=P_{\tau,1}^0\wedge P_{\tau,1}^1,\qquad A_{\tau,0}:=P_{\tau,0}^0 \lor P_{\tau,0}^1.
\]
\item Suppose the operation corresponding to $\tau$ is $\vee$. Similar to before, we can set
\[
A_{\tau,1}:=P_{\tau,1}^0\vee P_{\tau,1}^1,\qquad A_{\tau,0}:=P_{\tau,0}^0\wedge P_{\tau,0}^1.
\]
\item Suppose the operation corresponding to $\tau$ is the constant $1$. Then we set $A_{\tau,1} := \top$ and $A_{\tau,0} := \bot$. Similarly, if it is the constant $0$, we set $A_{\tau,1} := \bot$ and $A_{\tau,0} := \top$.

\item Suppose $\tau$ is an input gate. Then the value of $(\tau,u)$ is $v$ iff $u$ corresponds to a number $i$ with $1 \le i \le m$ and the $i^{th}$ input bit $x_i$ is $v$. Hence, $\sem{A_{\tau,v}}$
must contain all $u$ which encode a number $1 \le i \le m$ such that $x_i = v$. Therefore, for any number $1 \le i \le m$, if its binary representation is given by $i_1 i_2 \cdots i_{H}$, we simply set
\[
  A_{\tau,v}:=\bigvee_{1 \le i \le m,\ x_i=v} ([u_1 = i_1] \land [u_2=i_2]\wedge\dots\wedge[u_{H}=i_{H}]).
\]
\end{itemize}

This completes the construction of the desired relation program. We can now formally show its correctness.

\begin{lemma}\label{cr:thm:translation}
For every gate $(\tau,u)$ of $C$ and every $v\in\bits$, we have
\[
  u\in\sem{A_{\tau,v}}\quad\text{if and only if}\quad\mathrm{val}(\tau,u)=v .
\]
\end{lemma}
  
\begin{proof}
To begin with, we first recall the definition of the computation of $\sem V$ for any variable $V$: the set $X_V^j$ is the set of $V$ after $j$ rounds, we have $X_V^j \subseteq X_V^{j+1}$, and the fixed point of this computation is taken to be $\sem V$. Accordingly, we say that a string $u$ is in the set of $V$ after $j$ rounds if $u \in X_V^j$. With these definitions recalled, we now proceed to the proof.

To begin with, we first relate the variables $B_{\tau,v}^{p,i}$ to the runs of the program $f_{\tau,i}$. Let $p$ be a state of $f_{\tau,i}$ and let $u$ be a string such that $f_{\tau,i}$, started in $p$ on $u$ reaches a final state $\sigma$ with the string $u'$. 
We claim that
\begin{enumerate}[label=(\alph*)]
\item If $u'\in\sem{A_{\sigma,v}}$, then $u\in\sem{B_{\tau,v}^{p,i}}$;
\item If $u$ is in the set of $B_{\tau,v}^{p,i}$ after $j$ rounds, then $u'$ is in the set of $A_{\sigma,v}$ after
fewer than $j$ rounds.
\end{enumerate}
 
We prove both these claims by induction on the number of steps of the run from $(p,u)$. 
For the base case, if the number of steps is 0, then $p = \sigma, u' = u$ and both these
claims follow from the equation $B_{\tau,v}^{p,i}:=A_{p,v}$. 
For the induction step, suppose the number of steps is non-zero. Let
$(p'',u'')$ be the configuration of the Boolean program $f_{\tau,i}$ that is reached
from $(p,u)$ in one step. We do a case split:
\begin{itemize}
    \item Suppose the step from $(p,u)$ to $(p'',u'')$ was an instruction of the form
    $\textsf{set}(t,c;p'')$. By definition, $u'' = u\upd tc$. By construction, the equation
    for $B_{\tau,v}^{p,i}$ is $B_{\tau,v}^{p,i} := B_{\tau,v}^{p'',i}\langle t \leftarrow c\rangle$. This equation, the fact that $u'' = u \upd tc$, and the induction hypothesis applied to $B_{\tau,v}^{p'',i}$ together immediately imply that the induction step holds
    for $(p,u)$ as well.
    \item Suppose the step from $(p,u)$ to $(p'',u'')$ was an instruction of the form
    $\textsf{test}(t;p_0,p_1)$. By definition, $u'' = u$ and $p'' = p_0$ if $u_t = 0$
    and $p'' = p_1$ otherwise. By construction, the equation
    for $B_{\tau,v}^{p,i}$ is $B_{\tau,v}^{p,i} := ([u_t = 0] \land B_{\tau,v}^{p_0,i}) \lor ([u_t = 1] \land B_{\tau,v}^{p_1,i})$. This equation, and the induction hypothesis applied to $B_{\tau,v}^{p'',i}$ together immediately imply that the induction step holds
    for $(p,u)$ as well.
\end{itemize}
 
We now prove the main equivalence, namely for every gate $(\tau,u)$ and every $v\in\bits$, we have
\[
  u\in\sem{A_{\tau,v}}\quad\text{if and only if}\quad\mathrm{val}(\tau,u)=v .
\]
 
Let us first prove the right-to-left implication by induction on the structure of the circuit. 
For the base case, if $(\tau,u)$ is an input gate, then $\mathrm{val}(\tau,u) = v$ implies that $u$ must be the binary encoding of a number $1 \le i \le m$ such that $x_i = v$. The equation for the input gates given in Part 2 then immediately implies that $u \in \sem{A_{\tau,v}}$. If $(\tau,u)$ is a constant gate with value $v$, then the equation of $A_{\tau,v}$ is $A_{\tau,v} := \top$, and so $u \in \sem{A_{\tau,v}}$.
 
For the induction step, let $(\sigma_i,u_i)$ be the $i^{th}$ predecessor of the gate and $v_i$ its value. By the induction hypothesis, $u_i\in\sem{A_{\sigma_i,v_i}}$. By definition, $f_{\tau,i}$ starting from the state $\tau$ on $u$ in the Boolean program $f_{\tau,i}$ has the output $(\sigma_i,v_i)$. Therefore claim (a) gives us that $u\in\sem{P^{i}_{\tau,v_i}}$. 
Examining the equations of Part 2 for the operations $\lnot, \land$ and $\lor$ immediately allows
us to deduce that $u \in \sem{A_{\tau,v}}$ as well, thereby completing the induction step.

Now let us prove the left-to-right implication on the number of rounds taken by the fixed point computation. More precisely, we show by induction on the number of rounds $j$ that if 
$(\tau,u)$ is a gate and $u$ is in the set of $A_{\tau,v}$ after $j$ rounds, then $\mathrm{val}(\tau,u) = v$.
For the base case of $j = 0$, the claim is vacuously true, as no string can belong to the set of a variable after 0 rounds. For the induction step, first suppose $(\tau,u)$ is an input gate.
Then the equation of $A_{\tau,v}$ as given in Part 2 ensures that in any round $j \ge 1$, the assignment to $A_{\tau,v}$ contains exactly those encodings of numbers $1 \le i \le m$ with $x_i = v$.
It follows then that $\mathrm{val}(\tau,u) = v$. Next, suppose $(\tau,u)$ is a constant gate. Then one of $A_{\tau,0}, A_{\tau,1}$ has the equation $\top$ and the other has the equation $\bot$, and so $u$ can only be in the set of $A_{\tau,v}$ for the value $v$ of the gate.
 
On the other hand, suppose $(\tau,u)$ is either a $\lnot, \land$ or $\lor$ gate. We show the claim for when it is a $\lnot$ gate; the arguments for the other cases are analogous. By the construction of the equation for $A_{\tau,v}$, it follows that $u$ can be in the set of $A_{\tau,v}$ 
after $j$ rounds only if $u$ is in the set of $P_{\tau,(1-v)}^0$ after $j-1$ rounds. Let $(\sigma_0,u_0)$ be the ($0^{th}$) predecessor of $(\tau,u)$. By claim (b), it follows that $u_0$ is in the set of $A_{\sigma_0,(1-v)}$ after $j-1$ rounds. By induction hypothesis, this implies that the value of $(\sigma_0,u_0)$ is $1-v$ and so consequently the value of $(\tau,u)$ is $v$. The arguments for the other two cases are similar. This completes the induction step and thereby also the proof of the claim.
\end{proof}

We end with a final discussion on the size of the relation program. Note that the number of variables we have introduced is $O(S + mH)$, where $S$ is the size of the local description of the circuit, $m$ is the number of inputs and $H$ is its address length. Indeed, for every type and every state of each Boolean program, we have introduced a constant number of variables. Furthermore, there are $2H$ more variables of the form $[u_t = c]$ for $1 \le t \le H$ and $c \in \{0,1\}$.
Finally, each conjunction and disjunction appearing in the equation for the input gates requires a fresh variable, thereby giving rise to $O(mH)$ additional variables. It is then clear that this relation program can be constructed in polynomial time in $S+mH$.
 
We can now prove \cref{cr:thm:main}. Recall that the output gate of $C$ is of the form $(\mathsf{out},0^H)$. 
Let $R$ be the variable $A_{\mathsf{out},1}$. Applying \cref{cr:thm:translation} to this output gate, 
we get that the output of $C$ on $x$ is $1$ if and only if $0^H \in \sem{R}$, which proves~\cref{cr:thm:main}.



\section{From Relation Programs to Intersection Non-Emptiness of PDA and $(k-1)$ NFAs}\label{sec:int}
Let us fix a $k \ge 1$ for this section. We shall now convert relation programs to instances of the intersection non-emptiness problem for one PDA and $k-1$ many NFAs. Formally, we will prove the following theorem.
 
\begin{theorem}\label{rk:thm:instance}
Let $\Pi$ be a relation program over $H$ bits, let $A$ be one of its variables, and let $u\in\bits^H$. There are a PDA $M_1$ and NFAs $M_2,\dots,M_k$ such that:
\begin{itemize}
\item The intersection of the languages of $M_1,\dots,M_k$ is non-empty if and only if $u\in\sem A$;
\item The size of each $M_i$ is $O\big(|\Pi|\cdot\log|\Pi|\cdot H\cdot\log H\cdot2^{H/2k+c\sqrt{H}}\big)$, where $c$ is a constant that does not depend on $\Pi$, $H$ and $k$;
\item The input alphabet of each $M_i$ and the stack alphabet of $M_1$ have constant size;
\item The number of transitions of each $M_i$ is at most twice its number of states (in particular, the machines are sparse);
\item Each $M_i$ can be computed from $\Pi$, $A$ and $u$ in time $O(N\log N)$ on a multitape Turing machine, where $N$ is the maximum of the sizes of $M_1,\dots,M_k$.
\end{itemize}
\end{theorem}
 
We prove the theorem in three parts. In the first part, we build a single, ``large'' pushdown automaton (with non-constant input and stack alphabets) $P$ such that the language of $P$ is non-empty iff the given string $u$ belongs to the set $\sem A$. In the second part, we show how to split $P$ into $k$ ``small'' machines (one PDA and $(k-1)$ NFAs) such that the intersection of the languages of these machines is non-empty iff the language of $P$ is non-empty. However, these ``small'' machines will still have non-constant input and stack alphabets. So, in the final third part,
we show how to reduce the sizes of these alphabets to a constant. We proceed to describe the first part.
 
\subsection{Part 1: The Pushdown Automaton $P$}\label{rk:sec:machine}
 
Let $\Pi$ be a relation program over $H$ bits. Without loss of generality, we can assume that $H$ is a multiple of $2k$. (Otherwise we pad the program with more bits that no equation mentions, until we reach a multiple of $2k$). Let $A$ be a variable of $\Pi$ and let $u \in \{0,1\}^H$. In this part,
we construct a pushdown automaton $P$ such that the language of $P$ is non-empty
iff $u \in \sem{A}$.

To this end, we cut every string $w \in \{0,1\}^H$ into $2k$ \emph{blocks} of
$H/2k$ bits each and write it as
\[
  x_1\,x_2\cdots x_k\ \ y_1\,y_2\cdots y_k .
\]
We write $x$ for $x_1x_2\cdots x_k$, and similarly $y$; so the string $w$ is $x\,y$, and each of
$x,y$ has $H/2$ bits. For a block $x_j$, we write $x_j[i]$ for its $i^{th}$ bit, where
$1\le i\le H/2k$. Having set up this block definition, we now proceed to systematically describe the construction of $P$ in various steps.

\subsubsection*{Sub-Part I: The States of $P$} The states of the PDA $P$ will be pairs of the form $(\ell,r)$, where $\ell$ comes from a set of locations
and $r=r_1\cdots r_k$ is a string that consists of $k$ \emph{segments} $r_1,\dots,r_k\in\bits^W$. Here $W$ is a number slightly larger than $H/2k$ that we fix below. (Note that we use the term blocks to refer to strings in $\{0,1\}^{H/2k}$ and segments to refer to strings in $\{0,1\}^W$. We will use these two terms in exactly this meaning throughout).
Intuitively, the set of locations is a ``small'' set and so a state of $P$
remembers an element from this ``small'' set and a string $r$ where each segment $r_j$ is allowed to have slightly more bits than a block $x_j$.
 
For a segment $r_j$, we write $r_j[i]$ for its $i^{th}$ bit, where $1\le i\le W$. We do not distinguish between a segment $r_j$ and the number $\sum_{i=1}^{W} r_j[i]\,2^{i-1}$ that it represents in binary (so $r_j[1]$ is the least significant bit); we call this number the value of $r_j$. In the same way, we do not distinguish between a block $x_j$ and its value $\sum_{i=1}^{H/2k} x_j[i]\,2^{i-1}$, which is smaller than $2^{H/2k}$. 
 
Intuitively, the main reason we need the segments in the states to be longer than the blocks is that,
at some places in our construction, we will use the segments to store blocks as well as to perform
some specific basic arithmetic on the values of the blocks. This specific arithmetic that we will do requires some more bits in the control state. More precisely, in our construction
we will rely on an injective encoding of the blocks into segments with a very special property.
To be more specific, to each number $q < 2^{H/2k}$ (which is just the value of a block), we injectively assign a number $\varphi(q) < 2^W$ (which is just the value of a segment) such that the set $\{\varphi(q)\}$ contains no three-term arithmetic progression, i.e., $\varphi(q) + \varphi(q') = 2\varphi(q'')$ implies $q = q' = q''$. For constructing this mapping, we adapt the idea behind Behrend's construction~\cite{Behrend46}, which we now state here.

\paragraph*{3-term AP-free encoding.}
 
Let $\alpha = \big\lceil\sqrt{H/2k}\,\big\rceil$. Note that every number $q < 2^{H/2k} \le 2^{\alpha^2}$ can be written in base $2^\alpha$ uniquely as $q = \sum_{i=1}^{\alpha} q_i\,(2^{\alpha})^{i-1}$ with digits $q_1,q_2,\dots,q_{\alpha} < 2^\alpha$. From $q$, we now define a number $\varphi(q)$ as
\[
\varphi(q) \ =\ \sum_{i=1}^{\alpha} q_i\,(2^{\alpha+1})^{i-1} \ +\ \left(\sum_{i=1}^{\alpha} q_i^2\right) (2^{\alpha+1})^\alpha.
\]
From the equation, it might look like $q_1,q_2,\dots,q_{\alpha},\sum_{i=1}^{\alpha} q_i^2$ are the digits of $\varphi(q)$ in base $2^{\alpha+1}$. This is partially correct: $q_1,q_2,\dots,q_{\alpha}$ are indeed the lowest $\alpha$ digits of this number in base $2^{\alpha+1}$. However, $\sum_{i=1}^{\alpha} q_i^2$ can be almost as high as $\alpha 2^{2\alpha}$ and so, would not count as a ``digit'' in the strict sense. However, as far as intuition is concerned, we took the digits $q_1,\dots,q_{\alpha}$ of $q$ and we appended to this list the number $\sum_{i=1}^{\alpha} q_i^2$ and then found a way to encode this new list into a single number $\varphi(q)$. We shall now show that
$\varphi(q)$ has all of the nice properties that we had hinted at before.

\begin{lemma}\label{rk:lem:code}
The mapping $\varphi$ satisfies the following three properties:
 
\begin{enumerate}
\item \emph{Injective}: $\varphi$ is injective.
\item \emph{3-AP free}: If $\varphi(q)+\varphi(q') = 2\varphi(q'')$, then $q=q'=q''$.
\item \emph{Representable by $W$ bits}: $\varphi(q) < 2^{W-2}$ where $W = \alpha^2 + 4\alpha + 3$ (and so $W \in H/2k + O(\alpha)$). 
\end{enumerate}
\end{lemma}

\begin{proof}
Throughout the proof, for a number $q<2^{H/2k}$ with digits $q_1,\dots,q_{\alpha}$ as above, we write $L(q) = \sum_{i=1}^{\alpha} q_i (2^{\alpha+1})^{i-1}$ (lowest contribution) and $S(q) = \sum_{i=1}^{\alpha} q_i^2$ (square contribution), so that $\varphi(q) = L(q) + S(q)\,(2^{\alpha+1})^\alpha$. Since every digit satisfies $q_i < 2^\alpha < 2^{\alpha+1}$, the number $L(q)$ is written in base $2^{\alpha+1}$ with the digits $q_1,\dots,q_{\alpha}$, and in particular $L(q) < (2^{\alpha+1})^\alpha$.
 
\emph{Injective.} Since $\varphi(q) = L(q) +S(q) (2^{\alpha+1})^\alpha$, it follows that $L(q)$ is the remainder of $\varphi(q)$ modulo $(2^{\alpha+1})^\alpha$. Hence, if $\varphi(q) = \varphi(q')$ then $L(q) = L(q')$. By definition, the digits of $L(q)$ in base $2^{\alpha+1}$ are $q_1,\dots,q_\alpha$. So $L(q)=L(q')$ implies $q_i=q'_i$ for all $i$, i.e., $q=q'$. Hence $\varphi$ is injective.
 
\emph{3-AP free.} Let $q_i$, $q'_i$ and $q''_i$ be the digits of $q$, $q'$ and $q''$ respectively, for $1\le i\le\alpha$. Suppose $\varphi(q)+\varphi(q')=2\varphi(q'')$, i.e.,
\[
\sum_{i=1}^{\alpha}(q_i+q'_i)\,(2^{\alpha+1})^{i-1} \ +\ \big(S(q)+S(q')\big)(2^{\alpha+1})^\alpha \ =\ \sum_{i=1}^{\alpha}2q''_i\,(2^{\alpha+1})^{i-1} \ +\ 2S(q'')\,(2^{\alpha+1})^\alpha .
\]

Note that $q_i+q_i'$ and $2q_i''$ are the $i^{th}$ lowest digits of the left-hand side and right-hand side numbers under the base $2^{\alpha+1}$. It follows then that $q_i+q_i' = 2q_i''$ for every $i$. 
This immediately implies that $S(q)+S(q') = 2S(q'')$. Substituting $q''_i = (q_i+q'_i)/2$ into this equation, we get
\[
0 \ =\ S(q)+S(q')-2S(q'') \ =\ \sum_{i=1}^{\alpha}\Big(q_i^2+{q'_i}^2-\tfrac12(q_i+q'_i)^2\Big) \ =\ \frac12\sum_{i=1}^{\alpha}(q_i-q'_i)^2 .
\]
This implies that $q_i=q'_i$ for every $i$, and then also that $q''_i=q_i$. So $q=q'=q''$.
 
\emph{Representable by $W$ bits.} By construction, we have that $L(q) \le (2^{\alpha+1})^\alpha - 1$, and $S(q) \le \alpha(2^\alpha-1)^2 < \alpha\,2^{2\alpha} \le 2^{3\alpha}$ (using $\alpha\le2^\alpha$), so $S(q)\le 2^{3\alpha}-1$. Therefore
\[
\varphi(q) \ \le\ ((2^{\alpha+1})^\alpha-1) + (2^{3\alpha}-1)\,(2^{\alpha+1})^\alpha \ <\ 2^{3\alpha}(2^{\alpha+1})^\alpha \ =\ 2^{\alpha^2+4\alpha} \ <\ 2^{W-2}.
\]
Finally, $\alpha < \sqrt{H/2k}+1$ gives $\alpha^2 < H/2k + 2\sqrt{H/2k} + 1 \le H/2k + 2\alpha + 1$, and so $W < H/2k + 6\alpha + 4$, i.e., $W\in H/2k+O(\alpha)$.
\end{proof}

\paragraph*{Storing blocks in segments.} We say that a segment $r_j$ \emph{stores} a block $x_j$ if 
$r_j = \varphi(x_j)$. We say that a string $r = r_1\cdots r_k$ of $k$ segments \emph{stores} a string $x = x_1\cdots x_k \in \bits^{H/2}$ of $k$ blocks if $r_j$ stores $x_j$ for every $j$; we will denote this by $r = \varphi(x)$. Since $\varphi$ is injective, a segment stores at most one block. Note that not every segment stores a block.

We have now described what the states of our PDA look like, as well as the intuition behind them using the encoding $\varphi$. We now proceed to Sub-Part II, where we will specify the input letters of our PDA.

\subsubsection*{Sub-Part II: Input Letters and Notations for the Transitions}

We now describe the input letters of $P$ and the notation that we will use for its transitions. Each input letter of $P$ can be roughly thought of as a ``command'' that concerns one segment, or only the stack. More precisely, for every $1 \le j \le k$ (each segment), $1 \le i \le W$ (each position of the segment $r_j$) and $c \in \bits$, we will have
letters of the form
\[
\ins{test}(j,i,c),\quad \ins{set}(j,i,c),\quad
\ins{halve}(j),\quad \ins{put}(c),\quad \ins{take}(c).
\]
In addition, for every $1 \le j \le k$ (each block), $1 \le i \le H/2k$ (each position of the block stored in $r_j$) and $c\in\bits$, we will have letters of the form
\[
\ins{test\_blk}(j,i,c),\quad \ins{set\_blk}(j,i,c).
\]
Note that the first kind of letters works on the $W$ bits of a segment, and the second kind works on the $H/2k$ bits of the block that a segment stores.

The names of these letters probably already indicate the operations that they stand for. Formally, every letter $a$ decides which strings $r$ it can be read from, to which string $r'$ it leads, and what happens to the stack. For any two locations $\ell,\ell'$ and a letter $a$, we write
\[
  \ell\xrightarrow{\;a\;}\ell'
\]
for the \textbf{set of all transitions} that \textbf{from every} pair of the form $(\ell,r)$ with $r \in \{0,1\}^{kW}$, read the letter $a$ and move as follows.
\begin{itemize}
\item $a=\ins{test}(j,i,c)$: The notation $\ell \xrightarrow{a} \ell'$ denotes all transitions of the form
$(\ell,r)\xrightarrow{a\,\mid\,\nop}(\ell',r)$ for every $r$ with $r_j[i]=c$. Intuitively, this letter acts as a testing command and the PDA can continue on this letter iff the bit $r_j[i]$ is $c$.

\item $a=\ins{set}(j,i,c)$: The notation $\ell \xrightarrow{a} \ell'$ denotes all transitions of the form $(\ell,r)\xrightarrow{a\,\mid\,\nop}(\ell',r')$ where $r'$ is the same as $r$ except that $r'_j[i]=c$. So the bit $r_j[i]$ is set to $c$, irrespective of whatever it was before.

\item $a=\ins{halve}(j)$: The notation $\ell \xrightarrow{a} \ell'$ denotes all transitions of the form  $(\ell,r)\xrightarrow{a\,\mid\,\nop}(\ell',r')$ such that $r_j = 2\varphi(q)$ for some $q<2^{H/2k}$ and $r'$ is the same as $r$ except that $r'_j = \varphi(q)$. So the segment $r_j$ is halved, and the PDA can continue only if the result stores a block.

\item $a=\ins{put}(c)$: The notation $\ell \xrightarrow{a} \ell'$ denotes all transitions of the form $(\ell,r)\xrightarrow{a\,\mid\,\push\,c}(\ell',r)$ for every $r$. So the bit $c$ is pushed onto the stack.

\item $a=\ins{take}(c)$: The notation $\ell \xrightarrow{a} \ell'$ denotes all transitions of the form $(\ell,r)\xrightarrow{a\,\mid\,\pop\,c}(\ell',r)$ for every $r$. So the bit $c$ is popped from the stack, and the PDA can continue only if $c$ is the bit on top of the stack.

\item $a=\ins{test\_blk}(j,i,c)$: The notation $\ell \xrightarrow{a} \ell'$ denotes all transitions of the form
$(\ell,r)\xrightarrow{a\,\mid\,\nop}(\ell',r)$ such that $r_j$ stores a block whose $i^{th}$ bit is $c$, i.e., $r_j = \varphi(x_j)$ for some $x_j$ and $x_j[i] = c$. Intuitively, this letter tests the $i^{th}$ bit of the block stored in $r_j$, and the PDA can continue on this letter iff $r_j$ stores a block and this bit is $c$.

\item $a=\ins{set\_blk}(j,i,c)$: The notation $\ell \xrightarrow{a} \ell'$ denotes all transitions of the form $(\ell,r)\xrightarrow{a\,\mid\,\nop}(\ell',r')$ where $r_j = \varphi(x_j), r_j' = \varphi(x_j')$ for some $x_j, x_j'$
such that $x_j'$ is obtained from $x_j$ by setting its $i^{th}$ bit to $c$, and the rest of $r'$ is the same as $r$.
So the $i^{th}$ bit of the block stored in $r_j$ is set to $c$, and the PDA can continue only if $r_j$ stores a block. (Note that this may change many bits of the segment $r_j$).

\end{itemize}

There is one more letter, $\#$, which concerns neither $r$ nor the bits on the stack, and which will be read when the PDA calls or returns from a procedure. (We will expand upon the behavior of this letter in Sub-Parts III and IV).

\paragraph*{Abbreviations.} In our constructions, very often, we will read two of these letters one after the other, in order to move a bit between $r$ and the stack. For this we use the following abbreviations. In each of them, $\ell_0$ and $\ell_1$ are fresh locations that are used only in these transitions.
\begin{itemize}
\item $\ell\xrightarrow{\ins{push}(j,i)}\ell'$ stands for the transitions
\[
\ell\xrightarrow{\ins{test}(j,i,c)}\ell_c\xrightarrow{\ins{put}(c)}\ell' \qquad\text{for both bits $c$.}
\]
So the net-effect is to push the bit $r_j[i]$ onto the stack.
\item $\ell\xrightarrow{\ins{pop\_store}(j,i)}\ell'$ stands for the transitions
\[
\ell\xrightarrow{\ins{take}(c)}\ell_c\xrightarrow{\ins{set}(j,i,c)}\ell' \qquad\text{for both bits $c$.}
\]
So the net-effect is to pop the top-most bit from the stack and store it in  $r_j[i]$.
\item $\ell\xrightarrow{\ins{pop\_cmp}(j,i)}\ell'$ stands for the transitions
\[
\ell\xrightarrow{\ins{take}(c)}\ell_c\xrightarrow{\ins{test}(j,i,c)}\ell' \qquad\text{for both bits $c$.}
\]
So the net-effect is to pop the top-most bit from the stack, and the PDA can continue only if this bit equals $r_j[i]$.
\item $\ell\xrightarrow{\ins{guess\_blk}(j,i)}\ell'$ stands for the two sets of transitions given by 
\[
\ell\xrightarrow{\ins{set\_blk}(j,i,0)}\ell' \qquad \text{and} \qquad  \ell\xrightarrow{\ins{set\_blk}(j,i,1)}\ell' .
\]
So the net-effect is to set the $i^{th}$ bit of the block stored in $r_j$ to $0$ or to $1$, nondeterministically.
\end{itemize}

\begin{remark}
It may seem odd that we do not simply have letters that push the bit $r_j[i]$ or pop a bit into $r_j[i]$. The reason will become clear when we split $P$ into $k$ machines: then the machine that owns the stack and the machine that knows the segment $r_j$ are different, and so every letter must concern only one of them. The bit $c$ in the letters $\ins{test}(j,i,c)$ and $\ins{put}(c)$ tells both machines which bit is meant.
\end{remark}

This completes the description of the input letters and the notations for the transitions that we will employ in the next two sub-parts, which construct the actual transitions of $P$.

\subsubsection*{Sub-Part III: Procedures}

We now describe the transitions of $P$. We will do this by describing a PDA $P_A$ for each variable $A$ of the relation program $\Pi$.
$P$ would then be obtained by putting together all of these PDAs as one.

The PDA $P_A$ will be called the \emph{procedure} of $A$ (similar to procedures in programming languages, these procedures can be thought of as function blocks which are called by other function blocks). We describe these procedures in two sub-parts. In this sub-part, we describe what each procedure looks like and what its precise interpretation will be. In the next sub-part (Sub-Part IV), we then construct the procedures themselves.

To begin with we describe the locations, input alphabet and stack alphabet of the procedures.

\paragraph*{Locations, input and stack alphabets of $P_A$.}
A procedure $P_A$ will have the \emph{main locations}
\[
 [A,0],\ [A,1],\ \dots,\ [A,n_A]
\]
for some $n_A$. The first main location $[A,0]$ is called the \emph{entry} of the procedure, and the last main location $[A,n_A]$ is its \emph{exit}. The procedure is passed through in $n_A$ \emph{stages}, where stage $i$ leads from the main location $[A,i-1]$ to the main location $[A,i]$. In addition to the main locations, every stage $i$ will have its own fresh locations, which we call the \emph{inner locations of stage $i$}. Hence, states of $P_A$ will be of the form $(\ell,r)$ with $\ell$ a main or an inner location and $r \in \bits^{kW}$.
The input alphabet of $P_A$ will consist of all the input letters described in Sub-Part II. Finally, the stack alphabet of $P_A$ will have the letters $0, 1$
and one \emph{tag letter} $\tau_{B}$ for every variable $B$ of $\Pi$.

\paragraph*{Balanced runs and the ordering property.} Before we describe the transitions of each $P_A$, we set up some intuition and notation. In our construction, we will work extensively with a special class of runs, called \emph{balanced runs}, which we formally define.

\emph{Balanced runs.} Intuitively, a run is balanced if it ends with the same stack content $w$ with which it starts, and this $w$ is never touched during the run. Formally a run $(s_0,w_0),(s_1,w_1),\dots,(s_n,w_n)$ is called \emph{balanced} if $w_0 = w_n$ and each $w_t$ is of the form $w_0 v_t$ for some $v_t$. The \emph{level} of $(s_t,w_t) = (s_t,w_0 v_t)$ is the number of tag letters that appear in $v_t$ and the \emph{depth} of such a run is the largest level of its configurations. Note that there is a balanced run between $(s,w)$ and $(s',w)$ for some $w$ iff there is a balanced run between $(s,v)$ and $(s',v)$ for any $v$. For this reason,
we will say that there exists a balanced run from state $s$ to $s'$ to mean that
there exists a balanced run between $(s,w)$ and $(s',w)$ for some $w$.

\emph{The form of the procedures.} The procedure $P_A$ we construct (in Sub-Part IV) will have a very simple form. More precisely, every stage $i$ of $P_A$ will be of one of the following two kinds.
\begin{itemize}
\item \emph{Simple}: In this case, all the transitions belong to one of the kinds described in Sub-Part II (and hence never push or pop a tag letter). Furthermore, all the transitions start at either $[A,i-1]$ or at an inner location of stage $i$, and they end at an inner location of stage $i$ or at $[A,i]$.
\item \emph{Call}: In this case, the stage has no inner locations, and its transitions are of the form
\[
([A,i-1],r)\xrightarrow{\#\,\mid\,\push\,\tau_A}(e,r)\qquad\text{and}\qquad(e',r)\xrightarrow{\#\,\mid\,\pop\,\tau_A}([A,i],r)
\]
where $r$ is an arbitrary string in $\{0,1\}^{kW}$ and $e, e'$ are the entry and exit of some procedure $Q$. Furthermore, the last transition will be the only transition that pops $\tau_A$ at the exit of $Q$.
\end{itemize}
The procedure $P_A$ will have no other transitions. If we are guaranteed that our procedures will be of this kind, we can already prove the following \emph{ordering property}, which will be useful for our constructions.

\begin{quote}
\emph{The Ordering Property.} Suppose $\rho$ is a balanced run of depth $m$ which starts at a location of $P_A$. Then, the locations of all level-0 configurations in this run are locations of $P_A$. Furthermore, suppose the location of a level-0 configuration $c$ belongs to stage $i$ (i.e., the location of $c$ is either $[A,i-1]$ or an inner location of stage $i$). Let $c'$ be the next level-0 configuration after $c$ whose location is a main location. Then the location of $c'$ is $[A,i]$.
\end{quote}

Indeed, we can prove this property by double induction on the depth $m$ and the length of $\rho$. For the base case of $m = 0$, note that no tags can be pushed onto or popped from the stack. Hence, every transition must be a transition of a simple stage. Such a transition leads from $[A,i-1]$ or an inner location of stage $i$ to an inner location of stage $i$ or to $[A,i]$, from which the claim is clear by a further induction on the length of the run. For the induction step, consider the first transition of $\rho$. If it is a transition of a simple stage, then for the same reason as the base case, we are done by the secondary induction hypothesis on the rest of the run. Suppose this transition is not a transition of a simple stage. By definition, this transition cannot pop a tag, since $\rho$ is a balanced run. Hence, the only remaining case is that this first transition must push the tag $\tau_A$ into the stack and move to the entry of some procedure $Q$. This can only happen if the location of the starting configuration of $\rho$ is of the form $[A,i-1]$ for some $i$ such that stage $i$ is a call stage. Let $d$ be the configuration after this first transition.

Since $\rho$ is a balanced run, this tag must be popped from the stack at some point. Let $d', c'$ be the configurations just before and after popping this tag respectively. (Note that $c'$ is the first configuration after the beginning configuration of $\rho$ which is at level 0). It follows that the run between $d$ and $d'$ is a balanced run of depth at most $m-1$. Applying the induction hypothesis, we get that the location of $d'$ is a location of $Q$. Furthermore, by assumption on the structure of the transitions, a tag can be popped only at an exit of a procedure. Hence the location of $d'$ is an exit of $Q$. By assumption on the structure of the transitions, we immediately get that the location of $c'$ is $[A,i]$. Applying the secondary induction hypothesis on the shorter run from $c'$ now completes the proof of the induction.

\paragraph*{Balanced runs and the relation program.} Not only do balanced runs allow us to derive the above nice ordering property, they also allow us to have the following property, which relates them to the computation of the relation program $\Pi$. To state it, we recall how this computation proceeds in rounds.

\emph{Rounds.} Recall from \cref{cr:sec:relprog}, that for every variable $A$ and every $m\ge0$, we have a set $X^{m}_A\subseteq\bits^H$, the set of $A$ \emph{after $m$ rounds} given by:
\[
 X^{0}_A=\emptyset,\qquad X^{m+1}_A=F_A\big(X^{m}\big),
\]
where $F_A(X)$ is the set that the equation of $A$ produces from an assignment $X$ of sets to the variables. (For example, $F_A(X)=X_{A_1}\cap X_{A_2}$ if the equation is $A:=A_1\wedge A_2$). Finally $\sem A$ is the fixed point of this computation. Our construction of each $P_A$ will be such that we can equate each of these sets $X_A^m$ precisely to the balanced runs of $P_A$ of bounded depth, by the following Central Property. (Recall from Sub-Part I that a string $r\in\bits^{kW}$ of $k$ segments stores a string $x\in\bits^{H/2}$ of $k$ blocks if $r_j=\varphi(x_j)$ for every $j$).

\begin{quote}
\emph{The Central Property.} For every $m\ge0$, every variable $A$ and all $r,r'\in\bits^{kW}$, there is a balanced run of depth strictly less than $m$ from the state $([A,0],r)$ to the state $([A,n_A],r')$ if and only if $r$ stores some $x\in\bits^{H/2}$ and $r'$ stores some $y\in\bits^{H/2}$ such that $x\,y\in X^{m}_A$.
\end{quote}


With this intuition relating what the runs of our machine look like compared to the fixed point computation of $\Pi$, we are now in a position to precisely describe the transitions of each procedure $P_A$ and then prove that they indeed satisfy the Central Property, by induction on the depth $m$. This is done in the next sub-part.

\subsubsection*{Sub-Part IV: Constructing the Procedures}

We now define the procedure of a variable $A$, by a case analysis on the form of its equation. As we describe each case, we also explain how each one of them satisfies the Central Property, by induction on the depth $m$. Notice that irrespective of how the procedures are defined, there can be no balanced runs of depth strictly less than 0 and $X_A^0 = \emptyset$ for every variable $A$. Hence, the Central Property trivially holds for the base case of $m = 0$. We will now describe the procedure of each variable $A$ and show how they satisfy the Central Property for some $m+1$, assuming that we have already established the truth of the Central Property for $m$.

\newcommand{\case}[1]{\par\medskip\noindent\textbf{Case #1}\ }

\case{$A:=\top$} In this case, the procedure $P_A$ has the two main locations $[A,0]$ and $[A,1]$ and a single simple stage, which guesses every bit of every block, one after the other:
\[[A,0]\xrightarrow{\ins{guess\_blk}(1,1)}\ell_1\xrightarrow{\ins{guess\_blk}(1,2)}\ell_2\ \cdots\ \xrightarrow{\ins{guess\_blk}(k,H/2k)}[A,1].\]
Here $\ell_1,\ell_2,\dots$ are the inner locations of this stage. Recall that the letters of $\ins{guess\_blk}(j,i)$ can be read only if the segment $r_j$ stores a block, and that they replace it by a segment that stores a block with an arbitrary $i^{th}$ bit. Hence, from $([A,0],r)$ the procedure can reach $([A,1],r')$ iff $r$ stores some $x$ and $r'$ stores some $y$. It immediately follows that for any depth $m \ge 0$, there is a balanced run of depth strictly less than $m+1$ between $([A,0],r)$ and $([A,1],r')$ iff
$r$ stores some $x$ and $r'$ stores some $y$ with $x\, y \in X_A^{m+1} = \sem{\top}$, proving the Central Property for this case.

\case{$A:=\bot$} In this case, $P_A$ has the two main locations $[A,0]$ and $[A,1]$ and no transition. So it has no run from its entry to its exit. Hence, by the same reasoning as the previous case, this case also satisfies the Central Property.

\case{$A:=[u_t=c]$} In this case, $P_A$ has to intuitively ``accept'' all strings whose $t^{th}$ bit is $c$. To this end, the position $t$ of a string $u=x\,y$
is a position in some block $x_j$ or $y_j$; say it is the $i^{th}$ bit of that block. More precisely, $t$ is the $i^{th}$ position in block $x_j$ or $y_j$ respectively iff $t = (j-1)H/2k + i$ or
$t = H/2 + (j-1)H/2k + i$ respectively. Accordingly, the procedure $P_A$ is given by
\begin{align*}
&[A,0]\xrightarrow{\ins{test\_blk}(j,i,c)}[A,1]\xrightarrow{\ins{guess\_blk}(1,1)}\ell_1\ \cdots\ \xrightarrow{\ins{guess\_blk}(k,H/2k)}[A,2] &&\text{for bit $i$ of $x_j$,} \\
&[A,0]\xrightarrow{\ins{guess\_blk}(1,1)}\ell_1\ \cdots\ \xrightarrow{\ins{guess\_blk}(k,H/2k)}[A,1]\xrightarrow{\ins{test\_blk}(j,i,c)}[A,2] &&\text{for bit $i$ of $y_j$.}
\end{align*}
So $P_A$ has three main locations and two simple stages: one of these stages reads a single letter, and the other one guesses every bit of every block, similar to the construction for $\top$. In the first case, the required bit position is tested at the start, and then the state may end with any segment storing any $y$. In the second case it is the other way round: first an arbitrary $y$ is chosen, and then its required bit position is tested, so that only the runs that end with a segment storing a ``good'' $y$, i.e., $y$ such that $y_j[i] = c$ survive. Note that none of these transitions involve any stack operations.
By construction, it is immediately seen that for any depth $m \ge 0$, there is a balanced run of depth strictly less than $m+1$ between $([A,0],r)$ and $([A,2],r')$ iff $r$ stores some $x$ and $r'$ stores some $y$ with $x\, y \in X_A^{m+1} = \sem{[u_t = c]}$. This proves the Central Property for this case.

\case{$A:=A_1\vee A_2$} In this case, the procedure $P_A$ has the two main locations $[A,0]$ and $[A,1]$ and a single stage, which is a call. From $[A,0]$ it can jump to the entry of $P_{A_1}$ or to the entry of $P_{A_2}$ (by pushing the tag $\tau_A$), and from their exits it can come back to $[A,1]$ (by popping the tag $\tau_A$). So the transitions are, for all $r$,
\begin{align*}
&([A,0],r)\xrightarrow{\#\,\mid\,\push\,\tau_{A}}([A_1,0],r) &&\text{and}&& ([A_1,n_{A_1}],r)\xrightarrow{\#\,\mid\,\pop\,\tau_{A}}([A,1],r),\\
&([A,0],r)\xrightarrow{\#\,\mid\,\push\,\tau_{A}}([A_2,0],r) &&\text{and}&& ([A_2,n_{A_2}],r)\xrightarrow{\#\,\mid\,\pop\,\tau_{A}}([A,1],r).
\end{align*}

By construction, assuming the Central Property is true for depth $m$, it is immediately seen that the Central Property is true for depth $m+1$ in this case.

\case{$A:=A_1\langle t\leftarrow c\rangle$} Similar to the case of $[u_t = c]$, again let position $t$ be the $i^{th}$ bit in some block $x_j$ or $y_j$. We now do a further case split on these two cases.

\begin{itemize}
    \item Suppose $t$ is the $i^{th}$ bit in the block $x_j$. In this case, $P_A$ will have the following transitions for every $r \in \{0,1\}^{kW}$.
    \begin{gather*}
    [A,0]\xrightarrow{\ins{set\_blk}(j,i,c)}[A,1],\\
    ([A,1],r)\xrightarrow{\#\,\mid\,\push\,\tau_A}([A_1,0],r),\qquad
    ([A_1,n_{A_1}],r)\xrightarrow{\#\,\mid\,\pop\,\tau_A}([A,2],r).
    \end{gather*}
    Intuitively, $P_A$ sets the bit $x_j[i]$ to $c$ and then calls $P_{A_1}$. Once $P_{A_1}$ is done, it returns back to $P_A$. Let us see how this satisfies the Central Property. Suppose $r$ stores $x$ and $r'$ stores $y$ with $xy \in X_A^{m+1}$. Let $x'$ be the same as $x$ except the bit in the $t^{th}$ position is $c$, and let $\tilde r$ be the string that stores $x'$. Note that $x'y \in X_{A_1}^{m}$ and so by induction hypothesis there is a balanced run of depth strictly less than $m$ from $([A_1,0],\tilde r)$ to $([A_1,n_{A_1}],r')$. By construction it is easily seen that this can be transformed into a balanced run of depth strictly less than $m+1$ from $([A,0],r)$ to $([A,2],r')$, thereby giving the desired run.

     For the other direction, suppose we have a balanced run of depth strictly less than $m+1$ from $([A,0],r)$ to $([A,2],r')$. By the Ordering Property, there are level-0 configurations $(s_0,w_0), (s_1,w_1)$ and $(s_2,w_2)$
     (appearing in that order) such that the location of each $s_i$ is $[A,i]$. Note that the run from
     $(s_0,w_0)$ to $(s_1,w_1)$ must read the letter $\ins{set\_blk}(j,i,c)$. Hence the segment $r_j$ stores a block (say $x_j$), and $s_1 = ([A,1],\tilde r)$, where $\tilde r$ is the same as $r$ except that $\tilde r_j$ stores the block 
     $\tilde x_j$ obtained from $x_j$ by setting its $i^{th}$ bit to $c$. Notice that the run between $(s_1,w_1)$ and $(s_2,w_2)$ must be a push of $\tau_A$, followed by a balanced run of depth strictly less than $m$, followed by a pop of $\tau_A$. By
     construction and by the Ordering Property, this balanced run in between the pushes and the pops must be a run between
     $([A_1,0],\tilde r)$ and $([A_1,n_{A_1}],r')$. By induction hypothesis, $\tilde r$ stores some $x'$ and $r'$ stores some $y$ with $x'y \in X_{A_1}^m$. Since $\tilde r$ stores $x'$, by construction, the string $r$ stores some $x$, and $x'$ is the same as $x$ except that the bit in the $t^{th}$ position is $c$. This immediately implies that $xy \in X_{A}^{m+1}$, thereby proving the claim for this case.

    \item Suppose $t$ is the $i^{th}$ bit in the block $y_j$. In this case, $P_A$ will have the following transitions for every $r \in \{0,1\}^{kW}$.
    \begin{gather*}
    ([A,0],r)\xrightarrow{\#\,\mid\,\push\,\tau_A}([A_1,0],r),\qquad
    ([A_1,n_{A_1}],r)\xrightarrow{\#\,\mid\,\pop\,\tau_A}([A,1],r),\\
    [A,1]\xrightarrow{\ins{test\_blk}(j,i,c)}[A,2]\xrightarrow{\ins{guess\_blk}(j,i)}[A,3].
    \end{gather*}
    In this case, $P_A$ first calls $P_{A_1}$ from $([A,0],r)$. As a result of this,
    when the call is returned, $r$ stores some $x$ and $P_A$ is at a state of the form $([A,1],r')$ such
    that $r'$ stores some $y$ and $x\,y$ is in $X_{A_1}^{m}$. Now, $P_A$ checks if the $i^{th}$ bit of $y_j$ is $c$. If this is successful, then it immediately follows that the string $x\, y' \in X_A^{m+1}$ where $y'$ is the same as $y$ except the bit in the $t^{th}$ position can be either 0 or 1. Therefore, it guesses any bit for this position and moves to its exit. The proof of correctness of the Central Property is similar to the first case.
\end{itemize}

\case{$A:=A_1\wedge A_2$} Without loss of generality, let $A_1 \neq A_2$. This is the most tricky case and requires some non-trivial ideas. We would like to construct $P_A$ so that there is a balanced run of depth strictly less than $m+1$ between $([A,0],r)$ and $([A,n_A],r')$ iff $r$ stores some $x$ and $r'$ stores some $y$ with $xy \in X_A^{m+1}$, i.e., $xy \in X_{A_1}^{m}$ and $xy \in X_{A_2}^{m}$. Let us begin with a naive idea and then try to fix the issues that arise it in one-by-one to see how the final procedure looks like. 

A first attempt to construct $P_A$ would be to first call $P_{A_1}$ and then call $P_{A_2}$.
However, this will not be correct: $P_{A_1}$ would start from $([A_1,0],r)$, where $r = \varphi(x)$, and reach some $([A_1,n_{A_1}],r')$, where $r':=\varphi(y)$ such that $x\,y \in X_{A_1}^{m}$. Starting from here we cannot verify
if $x\,y \in X_{A_2}^{m}$, because $r = \varphi(x)$ is lost to us. 

\subparagraph{Fix 1: } To rectify this, before calling $P_{A_1}$, we push $r = \varphi(x)$ onto the stack. Then, when $P_{A_1}$ returns with some $r' = \varphi(y)$, we still have $r$ on the stack. We then replace $r'$ with $r$ on the state
and now run $P_{A_2}$ from here. This rectifies the previous problem, but it introduces a new problem. When $P_{A_2}$ ends, it returns with some $r'' = \varphi(y')$. We still have to check that $r'' = r'$, to conclude that $xy \in X_A^{m+1}$. However, $r'$ is lost to us. 

\subparagraph{Fix 2: } To rectify this, notice that when $P_{A_1}$ returns with $r'$, if we figure out a way to \emph{swap} the $r'$ on the state with the $r$ on the stack, then our problems are solved: We can now run $P_{A_2}$ from $r$ and when it ends with some $r''$ on the state, we can easily check $r'' = r'$, by popping $r'$ from the stack.
The problem with this idea is that we have no simple way of swapping $r$ with $r'$, as the state does not have
room for both $r$ and $r'$. 

\subparagraph{Fix 3: } However, the state does have room for the \emph{sum} of the segments of $r$ and $r'$. So after $P_{A_1}$ returns with some $r'$, using the copy of $r$ on the stack, we store in each segment $j$, the value $r_j+r'_j$ on the state. Then for each segment $j$, we guess a string $g_j$ bit-by-bit and push it onto the stack as well as simultaneously subtract it from the number $r_j+r'_j$ on the state. If we have correctly guessed each $g_j$ as $r'_j$, then we have successfully swapped $r$ and $r'$ between the state and the stack. Now, we just have to run $P_{A_2}$, end at some $r''$ and check that $r'' = g$. However, the correctness of this idea relies on as correctly guessing $g$ as $r'$.

\subparagraph{Fix 4: } We can penalize runs that do not correctly guess $g$ as $r'$ as follows (and this is where the magic happens). Suppose after checking $g = r''$, we had access to $r_j+r'_j+g_j$ and $r_j$ for every segment $j$. We can then replace each segment $r''_j$ in the state with $r_j+r'_j+g_j = r_j+r'_j+r''_j$ and then subtract $r_j$ from that to get $r'_j+r''_j$. To now check that $r'_j = g_j = r''_j$,
by\cref{rk:lem:code}, we just have to check if $r'_j+r''_j = 2\varphi(q_j)$ for some $q_j$. This can be done by using the $\ins{halve}(j)$ input letter that we have. Hence, if we had access to both $r+r'+g$ and $r$, then we are done.

\begin{figure}[t]
\centering
\resizebox{\textwidth}{!}{%
\begin{tikzpicture}[font=\small]
\newcommand{\seg}[2]{\draw[fill=black!6] (#1-1.3,2.15) rectangle ++(2.6,0.6); \node at (#1,2.45) {#2}; \draw[thick] (#1-1.1,0)--(#1+1.1,0);}
\newcommand{\cell}[3]{\draw (#1-1.0,#2*0.65+0.07) rectangle ++(2.0,0.58); \node at (#1,#2*0.65+0.36) {#3};}
\seg{0}{$r$}
\node[align=center] at (0,3.75) {$[A,0]$\\ \ \\ start};
\seg{2.9}{$r$}\cell{2.9}{0}{$r$}\cell{2.9}{1}{$r$}
\node[align=center] at (2.9,3.75) {$[A,1]$\\ Stage 1:\\ push twice};
\seg{5.8}{$r'$}\cell{5.8}{0}{$r$}\cell{5.8}{1}{$r$}
\node[align=center] at (5.8,3.75) {$[A,2]$\\ Stage 2:\\ call $P_{A_1}$};
\seg{8.7}{$s = r+r'$}\cell{8.7}{0}{$r$}
\node[align=center] at (8.7,3.75) {$[A,3]$\\ Stage 3:\\ add};
\seg{11.6}{$s-g$}\cell{11.6}{0}{$r$}\cell{11.6}{1}{$s+g,\ \ g$}
\node[align=center] at (11.6,3.75) {$[A,4]$\\ Stage 4:\\ guess $g$, split};
\seg{14.5}{$r''$}\cell{14.5}{0}{$r$}\cell{14.5}{1}{$s+g,\ \ g$}
\node[align=center] at (14.5,3.75) {$[A,5]$\\ Stage 5:\\ call $P_{A_2}$};
\seg{17.4}{$s+g$}\cell{17.4}{0}{$r$}
\node[align=center] at (17.4,3.75) {$[A,6]$\\ Stage 6:\\ compare, restore};
\seg{20.3}{$r'+r''$}
\node[align=center] at (20.3,3.75) {$[A,7]$\\ Stage 7:\\ subtract};
\seg{23.2}{$(r'+r'')/2$}
\node[align=center] at (23.2,3.75) {$[A,8]$\\ Stage 8:\\ halve};
\foreach \x in {1.45,4.35,7.25,10.15,13.05,15.95,18.85,21.75} {\draw[-{Stealth[length=1.8mm]}] (\x-0.14,2.45)--(\x+0.14,2.45);}
\node at (-2.1,0.36) {stack};
\node at (-2.1,2.45) {state};
\end{tikzpicture}%
}
\caption{A run of the procedure of $A:=A_1\wedge A_2$. Above, the string in the state at each main location; below, the contents of the stack. All sums and differences are taken segment by segment. The run starts with $r = \varphi(x)$. The call of $P_{A_1}$ returns $r' = \varphi(y)$, and $s$ denotes the sum $r+r'$. In Stage 4 a string $g$ is guessed, and the bits of $s+g$ and of $g$ are pushed in an interleaved manner. The call of $P_{A_2}$ is started from $s-g$ and returns $r''$, and Stage 6 can be completed only if $g=r''$; it then puts $s+g = r+r'+r''$ into the state. The last stage can be completed only if $r'_j+r''_j=2\varphi(q_j)$ for every segment $j$, which forces $r'=r''=g$, and hence $s-g=r$.}
\label{rk:fig:gadget}
\end{figure}
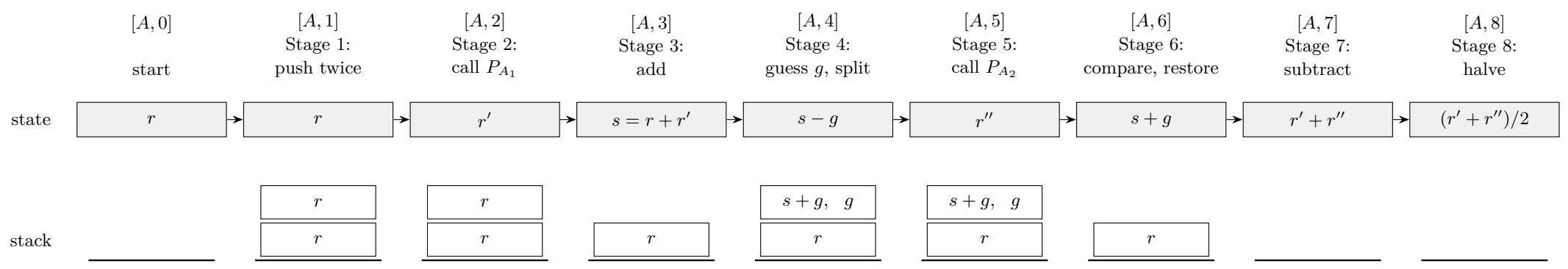

\subparagraph{Fix 5: } This can be easily accomplished as follows, and this brings us to the description of the complete idea (which is pictorially depicted in \cref{rk:fig:gadget}): We begin the procedure by initially pushing $r := \varphi(x)$ into the stack twice. Then we run $P_{A_1}$ from $r$ on the state and end with some $r' := \varphi(y)$ on the state. For each segment $j$, by popping $r_j$ from the stack and adding it to $r'_j$, we get $s_j := r_j+r'_j = \varphi(x_j) + \varphi(y_j)$ on the state (and still one copy left of $r$ on the stack). Then, for each segment $j$, we guess a string $g_j$ bit-by-bit, simultaneously subtracting it from $s_j$ (bit-by-bit) as well as pushing bit-by-bit the strings $s_j+g_j$ and $g_j$ in an interleaved manner onto the stack (so the first bit of $s_j+g_j$, followed by the first bit of $g_j$, followed by the second bit of $s_j+g_j$, followed by the second bit of $g_j$ and so on). Hence, at the end of this stage, for each segment $j$, we will be left with $s_j-g_j$ on the state and the stack will contain the bits of $s_j+g_j$ and $g_j$ in an interleaved manner, followed by a copy of $r$.

Now, from here, we run $P_{A_2}$ and end with some $r''$ on the state. For each segment $j$, we now pop the bits of $s_j+g_j$ and $g_j$ from the stack bit-by-bit, checking that each bit of $g_j$ popped corresponds with the bit of $r''_j$ in the same position, and if so replacing it with the corresponding bit of $s_j+g_j$. Hence, this successfully verifies that $g_j = r''_j$ for every segment $j$ and also ensures that the state now has $s_j+g_j = r_j+r'_j+g_j = r_j+r'_j+r''_j$ in the $j^{th}$ segment. Note that the stack still has a copy of $r$. Now, for each segment $j$, we subtract $r_j$ from $r_j+r'_j+r''_j$ to get $r'_j+r''_j$ on the state. We now check if $r'_j+r''_j = 2\varphi(q_j)$ for some $q_j$ by the $\ins{halve}(j)$ letters. By \cref{rk:lem:code}, this will ensure that $r'_j = r''_j$ for all $j$, giving us $r'  = r'' = g$. This implies that when we called $P_{A_1}$ and $P_{A_2}$, we called it from $r$ and $s-g = r$ and when they both returned, they returned with $r'$ and $r'' = r'$. By induction hypothesis, this means that $xy \in X_{A_1}^m \cap X_{A_2}^m$, thereby showing us that $xy \in X_{A}^{m+1}$. We now proceed to the formal details.

The procedure $P_A$ has the nine main locations $[A,0],[A,1],\dots,[A,8]$ and hence eight stages. Stages 2 and 5 are calls, and the other stages are simple. In the simple stages, all locations that are not main locations (written $\ell$ with some indices, or $\cdot$) are inner locations of that stage. We recommend that the reader consult the \cref{rk:fig:gadget} at the end of the description of each stage, to see how the state and the stack would look like.
 
\paragraph*{Stage 1: Push $r$ twice.} We begin by pushing all bits of $r$ onto the stack, segment by segment from $r_1$ to $r_k$, and within each segment from the most significant bit to the least significant bit. Then we do the same once more:
\begin{gather*}
[A,0]\xrightarrow{\ins{push}(1,W)}\ell_1\xrightarrow{\ins{push}(1,W-1)}\ell_2\ \cdots\ \xrightarrow{\ins{push}(1,1)}\ell_W\xrightarrow{\ins{push}(2,W)}\ell_{W+1}\ \cdots\ \xrightarrow{\ins{push}(k,1)}\ell_{kW},\\
\ell_{kW}\xrightarrow{\ins{push}(1,W)}\ell_{kW+1}\ \cdots\ \xrightarrow{\ins{push}(k,1)}[A,1].
\end{gather*}
Afterwards the stack contains two copies of $r$ on top, and each segment will later be popped from its least significant bit.
 
\paragraph*{Stage 2: Call $P_{A_1}$.} We now call $P_{A_1}$. More precisely, for every $r, r'\in\{0,1\}^{kW}$ we have
\[
([A,1],r)\xrightarrow{\#\,\mid\,\push\,\tau_{A}}([A_1,0],r),\qquad
([A_1,n_{A_1}],r')\xrightarrow{\#\,\mid\,\pop\,\tau_{A}}([A,2],r').
\]
 
\paragraph*{Stage 3: Add the upper copy.} We now pop the upper copy of the old $r$ from the stack and add it to the current $r'$ in the state, segment-by-segment in reverse order from $k$ to $1$ (and within each segment from the least significant bit). For this we use inner locations $\ell_{j,i,c}$, which mean ``the next bit to be treated is $i^{th}$ bit in the $j^{th}$ segment, and the current carry is $c$''. More precisely, for all $1\le j\le k$, $1\le i\le W$ and all bits $c,v,h$, let $o,c'$ be the bits with $v+h+c=o+2c'$. Then we have the transitions
\[
\ell_{j,i,c}\xrightarrow{\ins{test}(j,i,v)}\ \cdot\ \xrightarrow{\ins{take}(h)}\ \cdot\ \xrightarrow{\ins{set}(j,i,o)}\ell_{j,i+1,c'}.
\]
(The two locations $\cdot$ are fresh for every choice of $j,i,c,v,h$.) So the bit $v$ of the segment and the popped bit $h$ are added with the carry $c$, the result $o$ is written into the segment, and the new carry is $c'$. 

We have to move from the $k^{th}$ segment till the first segment and so this stage starts at $[A,2]:=\ell_{k,1,0}$.
Furthermore, after a segment is finished we have to move to the next one and so we set $\ell_{j,W+1,0}:=\ell_{j-1,1,0}$ for each $j > 1$. Finally the stage must end at the end of the first segment and so we set $[A,3] := \ell_{1,W+1,0}$.
 
\paragraph*{Stage 4: Guess and split.} For every segment $j$, we now guess a $g_j$ bit-by-bit and subtract it from the $j^{th}$ segment currently in the state ($s_j := r_j+r'_j$) to get a new segment ($s_j-g_j$). While doing this we also simultaneously add $g_j$ to the $j^{th}$ segment in the state and push this resulting sum ($s_j+g_j$) onto the stack, interleaved with pushing $g_j$ as well. For this purpose, we use inner locations $\ell_{j,i,c,d}$, where $c$ is the carry of the addition and $d$ is the borrow of the subtraction. More precisely, for all $1\le j\le k$, $1\le i\le W$ and all bits $c,d,v,g$, let $f,c',p,d'$ be the bits with $v+g+c=f+2c'$ and $v-g-d=p-2d'$. Then we have the transitions
\[
\ell_{j,i,c,d}\xrightarrow{\ins{test}(j,i,v)}\ \cdot\ \xrightarrow{\ins{set}(j,i,p)}\ \cdot\ \xrightarrow{\ins{put}(f)}\ \cdot\ \xrightarrow{\ins{put}(g)}\ell_{j,i+1,c',d'}.
\]
(The three locations $\cdot$ are fresh for every choice of $j,i,c,d,v,g$ and $g$ denotes the guessed bit).
This stage starts at $[A,3]:=\ell_{1,1,0,0}$. After a segment is finished we move to the next one and so we set $\ell_{j,W+1,0,0}:=\ell_{j+1,1,0,0}$. Finally the stage ends at $[A,4] := \ell_{k,W+1,0,0}$. Note that if we had correctly guessed $g_j = r'_j$ for each $j$, then we would have $r$ on the state and an interleaving of $s+r'$ and $r'$ on the stack.
 
\paragraph*{Stage 5: Call $P_{A_2}$.} We now call $P_{A_2}$. More precisely, for every $r,r''\in\{0,1\}^{kW}$ we have
\[
([A,4],r)\xrightarrow{\#\,\mid\,\push\,\tau_{A}}([A_2,0],r),\qquad
([A_2,n_{A_2}],r'')\xrightarrow{\#\,\mid\,\pop\,\tau_{A}}([A,5],r'').
\]
 
\paragraph*{Stage 6: Compare and restore.} We pop the pairs of bits $(s+g,g)$ pushed in Stage 4, segment by segment from $k$ to $1$ (and within each segment from the most significant bit). We compare each bit of $g$ that is popped with the corresponding bit of the string $r''$ currently in the state, and if they are equal, we replace that bit with the corresponding bit of $s+g$.
\begin{gather*}
[A,5]\xrightarrow{\ins{pop\_cmp}(k,W)}\ell_1\xrightarrow{\ins{pop\_store}(k,W)}\ell_2\xrightarrow{\ins{pop\_cmp}(k,W-1)}\ell_3\xrightarrow{\ins{pop\_store}(k,W-1)}\ell_4\ \cdots\\
\cdots\ \xrightarrow{\ins{pop\_cmp}(1,1)}\ell_{2kW-1}\xrightarrow{\ins{pop\_store}(1,1)}[A,6].
\end{gather*}
So the stage can be completed only if $g_j=r''_j$ for every $j$, and if so, the stage then replaces every $r''_j$ by $s_j+g_j$.
 
\paragraph*{Stage 7: Subtract the lower copy.} This stage is the same as Stage 3, from $[A,6]$ to $[A,7]$, except that $o,c'$ are the bits with $v-h-c=o-2c'$. So $c$ is now a borrow, the lower copy of the old $r$ is popped and subtracted from the current $s+g = s+r''$, and at the end of this stage, the $j^{th}$ segment of the state will contain $r'_j+r''_j$.
 
\paragraph*{Stage 8: Halve.} We now halve every segment. More precisely, we have the following transitions:
\[
[A,7]\xrightarrow{\ins{halve}(1)}\ell_1\xrightarrow{\ins{halve}(2)}\ell_2\ \cdots\ \xrightarrow{\ins{halve}(k)}[A,8].
\]
Recall that this is possible only if the current value in every segment $j$ ($r'_j+r''_j$) is of the form $2\varphi(q_j)$ for some $q_j$. By \cref{rk:lem:code}, it then follows that $r'_j = r''_j = g_j$, which gives the correctness of the gadget.
 
Let us now show that this procedure satisfies the Central Property for some $m+1$. Suppose we have a balanced run of depth strictly less than $m+1$ that starts in $([A,0],r)$ and ends in $([A,8],r')$. By the Ordering Property, the run contains level-0 configurations $(s_0,w_0),(s_1,w_1),\dots,(s_{8},w_{8})$, in this order, such that the location of each $s_i$ is $[A,i]$. Similar to the previous case, by construction, the run
after the push of $(s_1,w_1)$ and before the pop to $(s_2,w_2)$ is a balanced run of depth strictly less than $m$ from the entry to the exit of $P_{A_1}$, which starts with the string of $s_1$ and ends with the string of $s_2$. Similarly, the run after the push of $(s_4,w_4)$ and before the pop to $(s_5,w_5)$ is a balanced run of depth strictly less than $m$ from the entry to the exit of $P_{A_2}$, which starts with the string of $s_4$ and ends with the string of $s_5$.
 
Tracing the construction of $P_A$, let us now see how each $s_i$ and $w_i$ look like. We use the following notation. For two strings $a,b$ of $k$ segments, we write $a+b$ (and $a-b$) for the string whose $j^{th}$ segment is $a_j+b_j$ (respectively $a_j-b_j$), for every $j$. For a segment $a_j$, we write $\overleftarrow{a_j}$ for the word $a_j[W]\,a_j[W-1]\cdots a_j[1]$, and for a string $a$ of $k$ segments, we write $\overleftarrow{a}$ for the word $\overleftarrow{a_1}\,\overleftarrow{a_2}\cdots\overleftarrow{a_k}$. Finally, for two segments $a_j,b_j$, we write $a_j\bowtie b_j$ for their perfect shuffle, i.e., the word $a_j[1]\,b_j[1]\,a_j[2]\,b_j[2]\cdots a_j[W]\,b_j[W]$. (We always denote the topmost entry of the stack as the rightmost one).
 
\begin{enumerate}[label=\arabic*.]
\item By assumption, $s_0 = ([A,0],r)$, and $w_0$ is some stack content.
\item By construction, $s_1 = ([A,1],r)$ and $w_1 = w_0\,\overleftarrow{r}\,\overleftarrow{r}$.
\item Let $r'$ be the string of $s_2$. By induction hypothesis for the call of $P_{A_1}$, we have $r = \varphi(x)$ for some $x$ and $r' = \varphi(y)$ for some $y$ with $xy \in X_{A_1}^m$. So $s_2 = ([A,2],r')$, and $w_2 = w_1$.
\item By construction, $s_3 = ([A,3],s)$ with $s := r+r'$, and $w_3 = w_0\,\overleftarrow{r}$.
\item By construction, $s_4 = ([A,4],s-g)$ for some string $g$ of $k$ segments, and $w_4 = w_0\,\overleftarrow{r}\,(f_1\bowtie g_1)\,(f_2 \bowtie g_2)\cdots(f_k \bowtie g_k)$, where $f := s+g$.
\item Let $r''$ be the string of $s_5$. By induction hypothesis for the call of $P_{A_2}$, we have $s-g = \varphi(x')$ for some $x'$ and $r'' = \varphi(y')$ for some $y'$ with $x'y'\in X_{A_2}^m$. So $s_5 = ([A,5],r'')$, and $w_5 = w_4$.
\item By construction, the run can move from $s_5$ to $s_6$ only if $g = r''$. Then $s_6 = ([A,6],f)$ with $f = s+g = r+r'+r''$, and $w_6 = w_0\,\overleftarrow{r}$.
\item By construction, $s_7 = ([A,7],f-r)$ with $f-r = r'+r''$, and $w_7 = w_0$.
\item By construction, the run can move from $s_7$ to $s_8$ only if for every $j$, we have $r'_j+r''_j = 2\varphi(q_j)$ for some $q_j<2^{H/2k}$. Then $s_8 = ([A,8],\varphi(q))$ with $q := q_1\cdots q_k$, and $w_8 = w_0$.
\end{enumerate}
 
For every $j$, we have $\varphi(y_j)+\varphi(y'_j) = r'_j+r''_j = 2\varphi(q_j)$, and so by the second property of \cref{rk:lem:code}, $y_j = y'_j = q_j$. Hence $y = y' = q$, and the string of $s_8$ is $\varphi(q) = \varphi(y)$. Moreover $g = r'' = \varphi(y') = \varphi(y) = r'$, hence $\varphi(x') = s-g = r$, and so $x' = x$ because $\varphi$ is injective. Therefore, $r$ stores $x$, the string of $s_8$ stores $y$, and $xy \in X_{A_1}^{m}\cap X_{A_2}^{m}$, implying that $xy \in X_{A}^{m+1}$. This shows one direction of the Central Property for $m+1$ for this case.

For the other direction, suppose $r$ stores $x$ and $r'$ stores $y$ with $xy\in X_A^{m+1}$, so that $xy \in X_{A_1}^{m}$ and $xy \in X_{A_2}^{m}$. We take the run from $([A,0],r)$ that follows the construction, uses the runs given by the induction hypothesis for the two calls for $P_{A_1}$ and $P_{A_2}$, and guesses $g_j = \varphi(y_j)$ for every $j$. It can then be easily verified that the run ends in $([A,8],r')$. This allows us to conclude the proof.

\paragraph*{Wrapping up.}
Now, let $\Pi$ be a relation program, $A$ a variable and $u=x\,y$ a string. Let $P$ consist of the procedures $P_{B}$ of all variables $B$ of $\Pi$. Let $r$ be the string that stores $x$ and let $r'$ be the string that stores $y$. By the Central Property, it follows that starting and ending with the empty stack, there is a run from $([A,0],r)$ to $([A,n_A],r')$ iff $xy \in \sem{A}$. Hence, if we take these to be the initial and final states of $P$, then the language of $P$ is non-empty iff a given string $u \in \sem{A}$ for a given variable $A$. In the next part, we show that this $P$ can be split into $k$ ``small'' machines.
In the part after that, we will show how to reduce the input and stack alphabets so that they are constant-sized.

\subsection{Part 2: Splitting $P$ Into $k$ Machines}\label{rk:sec:split}

We now show how to split the ``large'' PDA $P$ (whose number of states is $O(|\Pi| \cdot H \cdot 2^{kW})$)
into one ``small'' PDA and $k-1$ many ``small'' NFAs (whose number of states is $O(|\Pi| \cdot H \cdot 2^{W})$).
As we shall see, this is quite simple, precisely because of the way we chose our input letters: every letter concerns either one segment of the state or the stack, but never both, and never two segments. Indeed,
intuitively speaking, the new PDA will store the location and the segment $r_1$ of the states
and the $j^{th}$ NFA (for $j$ between 2 and $k$) will store the segment $r_j$ of the states.

More formally, for $j=2,\dots,k$, we construct the NFA $M_j'$ as follows. 
Its states are the segments $r_j\in\bits^{W}$. Furthermore, for every $i$ and $c\in\bits$, from a state $r_j$, it has the following transitions:
\begin{itemize}
\item On the letter $\ins{test}(j,i,c)$, it stays in the state $r_j$, and it can do so only if $r_j[i]=c$.
\item On the letter $\ins{set}(j,i,c)$, it moves from $r_j$ to $r_j'$, where $r_j'$ is the same as $r_j$ except that $r_j'[i] = c$.
\item On the letter $\ins{halve}(j)$, it moves from $r_j$ to $r_j'$, and it can do so only if $r_j = 2\varphi(q)$ and $r_j' = \varphi(q)$ for some $q < 2^{H/2k}$.
\item On the letter $\ins{test\_blk}(j,i,c)$, it stays in the state $r_j$, and it can do so only if $r_j$ stores a block whose $i^{th}$ bit is $c$.
\item On the letter $\ins{set\_blk}(j,i,c)$, it moves from $r_j$ to $r_j'$, and it can do so only if $r_j$ stores a block $x_j$ and $r_j'$ stores the block obtained from $x_j$ by setting its $i^{th}$ bit to $c$.
\item On every other letter (that is, on the letters about the other segments, and on $\ins{put}(c)$, $\ins{take}(c)$ and $\#$), it simply stays at $r_j$.
\end{itemize}
In other words, on the letters about the segment $r_j$, the NFA $M_j'$ does to its state exactly what these letters do to the segment $r_j$ in $P$ (as described in Sub-Part II of Part 1).

The PDA $M_1'$ is the same as $P$, except that it has forgotten all segments other than $r_1$. 
More precisely, its states are the pairs $(\ell,r_1)$, where $\ell$ is a location of $P$ and $r_1\in\bits^{W}$. Furthermore, it has the following transitions:
\begin{align*}
&\text{For every } (\ell,r)\xrightarrow{a\,\mid\,\mathit{op}}(\ell',r') \text{ in } P
 &&\text{there is}\quad (\ell,r_1)\xrightarrow{a\,\mid\,\mathit{op}}(\ell',r_1') \text{ in } M_1'.
\end{align*}

By construction, it then easily follows that for any letter $a$, for any $\ell, \ell'$, for any $r_1,\dots,r_k,r_1',\dots,r_k'$ and for any stack contents $w,w'$, there is a step of the form
$((\ell,r_1),w) \xrightarrow{a} ((\ell',r_1'),w')$ in $M_1'$ and steps of the form $r_j \xrightarrow{a} r_j'$ in each $M_j'$ iff there is a step of the following form in $P$
\[((\ell,r_1r_2\cdots r_k),w) \xrightarrow{a} ((\ell',r_1'r_2'\cdots r_k'),w').\]

Let $(\ell,r)$ and $(\ell',r')$ be the initial and final states of $P$. We then set $(\ell,r_1), r_2,$
$\dots,r_k$ and $(\ell',r_1'),r_2',\dots,r_k'$ to be the initial and final states of $M_1',\dots,M_k'$. It then follows that the intersection of the languages of these $k$ machines is non-empty iff the language of $P$ is non-empty.

By construction, it is clear that each $M_j'$ for $2 \le j \le k$ has $2^{W}$ states and $O(kW) = O(H)$ input letters, because $k$ is a constant. It can also be easily verified that there is at most one outgoing transition from a state for a letter. Hence, each $M_j'$ for $2 \le j \le k$ has size $O(H\cdot2^{W})$. Similarly, we can show that the number of states of the PDA $M_1'$ is $O(|\Pi| \cdot H\cdot2^{W})$: each procedure $P_A$ has at most eight stages, and each stage has $O(kW) = O(H)$ inner locations. Hence each procedure $P_A$ has $O(H)$ locations and therefore $O(H\cdot2^{W})$ states in $M_1'$. Combined over all possible procedures we get that the number
of states of $M_1'$ is $O(|\Pi| \cdot H\cdot2^{W})$. 

Coming to the number of transitions of $M_1'$, we first notice that from every state $q$, the number of input letters for which there is an outgoing transition from $q$ is at most two. Furthermore, if any of those letters $a$ is not equal to $\#$, then the number of outgoing transitions labelled by $a$ from $q$ is also at most two. Hence, the number of transitions that are not labelled by the input letter $\#$ is $O(|\Pi| \cdot H\cdot2^{W})$. Finally, note that whenever $\#$ is read, a tag is pushed onto or popped from the stack. Since each tag is pushed or popped by only a constant number of transitions for each $r_1$, it follows that the total number of transitions labelled by $\#$ is 
$O(|\Pi| \cdot2^{W})$. Hence, the overall size of $M_1'$ is $O(|\Pi| \cdot H\cdot2^{W})$.

Finally, by \cref{rk:lem:code} we have $W < H/2k+6\alpha+4$ and $\alpha < \sqrt{H/2k}+1$, and hence $2^{W} = O\big(2^{H/2k+6\sqrt{H/2k}}\big) = O\big(2^{H/2k+5\sqrt{H}}\big)$. So the intersection of the languages of these machines is non-empty iff $u \in \sem A$, and they have the size claimed in \cref{rk:thm:instance}. However, their input alphabet has $O(H)$ letters and the stack alphabet of the PDA has $O(|\Pi|)$ letters. This is what we will rectify in the next part, which will also make all the machines sparse.

\subsection{Part 3: Alphabets of Constant Size}\label{rk:sec:alphabets}

In this final part we replace both the input and stack alphabets of the machines $M_1',\dots,M_k'$ by alphabets of constant size. The idea is simple: The stack alphabet consists of $0, 1$ and a tag $\tau_A$ for every variable $A$ and so we simply replace each tag $\tau_A$ by a unique word in $\{a,b\}^{\lceil\log |\Pi|\rceil}$. Similarly, we replace each
input letter by encoding it in binary over $\{0,1\}$. 

Formally, we begin by treating the stack alphabet first. To this end, let $\sigma=\lceil\log|\Pi|\rceil$. 
We take two new stack letters $a$ and $b$, and we choose for every tag $\tau_A$, a
unique word $c(\tau_A)\in\{a,b\}^\sigma$. We then notice that every push and pop of a tag occurs only when a $\#$ is read by the PDA. Furthermore, each of the NFAs, upon reading $\#$, simply stays at its current state. It follows that
we can replace each push transition $\delta$ of the form $s\xrightarrow{\#\,\mid\,\push\,\tau_A}s'$
with 
\[
s\xrightarrow{\#\,\mid\,\push\,c_1}[\delta,1]\xrightarrow{\#\,\mid\,\push\,c_2}[\delta,2]\ \cdots\ [\delta,\sigma-1]\xrightarrow{\#\,\mid\,\push\,c_\sigma}s' ,
\]
where $[\delta,1],\dots,[\delta,\sigma-1]$ are fresh states and $c_i$ is the $i^{th}$ letter of $c(\tau_A)$.
Similarly, we can replace each pop transition $\delta$ of the form $s\xrightarrow{\#\,\mid\,\pop\,\tau_A}s'$ by
\[
s\xrightarrow{\#\,\mid\,\pop\,c_\sigma}[\delta,1]\xrightarrow{\#\,\mid\,\pop\,c_{\sigma-1}}[\delta,2]\ \cdots\ [\delta,\sigma-1]\xrightarrow{\#\,\mid\,\pop\,c_1}s' .
\]
Let this new machine be $M_1''$. Because of the property that all the NFAs, upon reading $\#$, simply stay at their current state, it follows that this replacement preserves the intersection. More precisely, because of this property it follows that the intersection of the languages of $M_1',\dots,M_k'$ is non-empty iff the intersection of the languages
of $M_1'',M_2',\dots,M_k'$ is non-empty. Since the number of transitions in $M_1'$ was $O(|\Pi|\cdot H\cdot2^{W})$, it follows that the size of $M_1''$ is $O(|\Pi|\log |\Pi| \cdot H\cdot2^{W})$.

We now change the input alphabet of the machines as follows. Recall that the size of the input alphabet is some $e = O(H)$.
Let $\lambda = \lceil \log e \rceil$. Now, we encode each input letter $d$ in binary using some unique word $c(d) \in \{0,1\}^{\lambda}$. Then we replace each transition $\delta$ of the form $s \xrightarrow{d} s'$ of the NFAs by
\[
s\xrightarrow{\;c_1\;}[\delta,1]\xrightarrow{\;c_2\;}[\delta,2]\ \cdots\ [\delta,\lambda-1]\xrightarrow{\;c_\lambda\;}s' ,
\]
where $[\delta,1],\dots,[\delta,\lambda-1]$ are fresh states and $c_i$ is the $i^{th}$ letter of $c(d)$. 
Similarly, we replace each transition $\delta$ of the form $s \xrightarrow{d\, \mid\, \mathit{op}} s'$ of the PDA by
\[
s\xrightarrow{c_1\,\mid\,\nop}[\delta,1]\xrightarrow{c_2\,\mid\,\nop}[\delta,2]\ \cdots\ [\delta,\lambda-1]\xrightarrow{c_\lambda\,\mid\,\mathit{op}}s' ,
\]
where only the last transition performs the operation $\mathit{op}$ on the stack. Let these new machines be $M_1,\dots,M_k$.
It is clear that the intersection of the languages of $M_1,\dots,M_k$ is non-empty iff the intersection of the languages
of $M_1'',M_2',\dots,M_k'$ is non-empty. Note that the size of each $M_i$ is $\lambda$ times the size of $M_i'$ (or of $M_1''$ in the case of $M_1$). Hence, we conclude that each $M_i$ has size $O(|\Pi|\log |\Pi| \cdot H \log H\cdot2^{W})$, which is $O(|\Pi|\log |\Pi| \cdot H \log H\cdot2^{H/2k+5\sqrt H})$ as we have seen at the end of Part 2. Finally, note that the machines $M_1,\dots,M_k$ are sparse. Indeed, in this last step, a machine with $r$ states and $t$ transitions is turned into a machine with $r+(\lambda-1)t$ states and $\lambda t$ transitions, and $\lambda t \le 2(r+(\lambda-1)t)$ because $\lambda \ge 2$.

Now, it is clear that $M_1,\dots,M_k$ satisfy the first four properties of Theorem~\ref{rk:thm:instance}. For the last item, let $N$ be the maximum of the sizes of the machines. Every state and every transition of a machine is described by a constant number of the following pieces: a location of a procedure, a segment, an input letter, or one of the newly introduced states to reduce the input and the stack alphabets. Each of them can be written with $O(\log N)$ bits. Moreover, for all letters other than $\ins{halve}(j)$, $\ins{test\_blk}(j,i,c)$ and $\ins{set\_blk}(j,i,c)$, the transitions that start at a state are obtained from the description of this state by changing a constant number of bits, by either increasing the index $i$ of a fresh state $[\delta,i]$, or by moving to the next location of the procedure, or by replacing the location by the entry or the exit of another procedure (the latter can be read off from the equation of the variable). So a multitape Turing machine can go through the equations of $\Pi$ one after the other and, for every location of the procedure of an equation, run a counter through all segments $r_1$ and write down the transitions of the PDA that start there; similarly for the NFAs as well. Since writing down one transition takes time $O(\log N)$, the total time for these transitions is $O(N\log N)$.

It remains to consider the three kinds of letters that work on the block that a segment stores. Their transitions start only at segments of the form $\varphi(q)$ or $2\varphi(q)$ for a number $q<2^{H/2k}$. So, for every location (in the case of the PDA) and every such letter, the Turing machine runs a counter through all numbers $q<2^{H/2k}$ instead, computes $\varphi(q)$ (and, for $\ins{set\_blk}(j,i,c)$, also $\varphi(q')$ for the number $q'$ obtained from $q$ by setting its $i^{th}$ bit to $c$) directly from the definition of $\varphi$, and writes down the transition. Computing $\varphi(q)$ takes time polynomial in $H$. Since there are only $2^{H/2k}$ such numbers $q$, whereas the machines have $2^{W}$ states for every location, and since $2^{W-H/2k} \ge 2^{4\alpha}$ grows faster than every polynomial in $H$, the total time for these transitions is $O(N)$. Hence all machines can be computed in time $O(N\log N)$. This proves Theorem~\ref{rk:thm:instance}.

\section{Putting It All Together}\label{sec:fin}
In this section, we will combine the three reductions from the previous three sections to prove our final result, namely \cref{fin:thm:main}.
 
Suppose, for the sake of contradiction, that some deterministic multitape Turing machine $\mathcal S$ solves the intersection non-emptiness problem for one PDA and $k-1$ NFAs in time $O(N^{2k-\delta})$ for some $\delta > 0$ and $k \ge 1$, where $N$ is the maximum size of the given machines. (In fact, it even suffices to suppose that $\mathcal S$ does this only for sparse machines whose input and stack alphabets are of constant size). Without loss of generality, we may assume that $\delta\le\frac12$. Let $d\ge1$ be a number such that the relation program of \cref{cr:thm:main} can be computed in time $O((S+mH)^d)$, and let
\[
e\ =\ \max\Big\{\Big\lceil\frac{9k^2}{\delta}\Big\rceil,\ d+1\Big\}.
\]
By the time hierarchy theorem, there is a language $L$ that is decided by some deterministic multitape Turing machine in time $T(m)=O(m^e)$, but by no such machine in time $O(m^{e-\delta/2})$. We will show that $L$ can be decided by a multitape Turing machine in time $O(m^{e-\delta/2})$, which is a contradiction.

\paragraph*{The multitape Turing machine.} Let $\xi$ be an input of length $m$, and let $T=T(m) = O(m^e)$.
\begin{enumerate}
\item By \cref{tm:thm:main}, we compute a local description of a circuit $C$ with address length $H=\log T+O(1)$ and size $S$ polynomial in $H$, such that $\xi\in L$ iff the output of $C$ on $\xi$ is $1$. 
\item By \cref{cr:thm:main}, we compute a relation program $\Pi$ over $H$ bits of size $O(S+mH)$ and a variable $R$ such that the output of $C$ on $\xi$ is $1$ iff $0^H\in\sem R$.
\item By \cref{rk:thm:instance}, we compute a PDA $M_1$ and NFAs $M_2,\dots,M_k$ such that $0^H\in\sem R$ iff the intersection of their languages is non-empty. These machines are sparse, and their input and stack alphabets are of constant size.
\item We run $\mathcal S$ on $M_1,\dots,M_k$ and give the same answer.
\end{enumerate}
The correctness of this algorithm follows from \cref{tm:thm:main,cr:thm:main,rk:thm:instance}.

\paragraph*{The running time.} We have $H=O(\log m)$ and $2^H=O(T)=O(m^e)$. Hence $S$ is polynomial in $\log m$, and $|\Pi|=O(S+mH)=O(m\log m)$. Let $N$ be the maximum of the sizes of $M_1,\dots,M_k$. By \cref{rk:thm:instance}, for some constant $c$,
\[
N\ =\ O\big(|\Pi|\cdot\log|\Pi|\cdot H\cdot\log H\cdot2^{H/2k+c\sqrt H}\big)\ =\ O\big(m^{2+\frac{e}{2k}}\big).
\]
Indeed, $2^{H/2k} = O(m^{\frac{e}{2k}})$ because $2^H = O(m^e)$. Moreover, $|\Pi|\cdot\log|\Pi|\cdot H\cdot\log H = O(m\log^4m)$, and $2^{c\sqrt H} = 2^{O(\sqrt{\log m})}$ because $H = O(\log m)$. Both $\log^4 m$ and $2^{O(\sqrt{\log m})}$ are $O(m^{1/2})$, and so the product of all the factors other than $2^{H/2k}$ is $O(m^2)$.
We now bound the time of each step of the above multitape Turing machine that we had described.
\begin{itemize}
\item Step 1 takes time polynomial in $H$, that is, polynomial in $\log m$.
\item Step 2 takes time $O((S+mH)^d)=O(m^d\log^dm)$. As $d\le e-1$, this is $O(m^{e-\delta/2})$.
\item Step 3 takes time $O(N\log N)=O(m^{3+\frac{e}{2k}})$. We have $3+\frac{e}{2k}\le e-\delta$: Indeed, showing this is the same as showing $e\cdot(1-\frac1{2k})\ge3+\delta$, which holds because $1-\frac1{2k}\ge\frac12$ and $e\ge\frac{9k^2}{\delta}\ge18$ and $\delta\le\frac12$. So Step 3 takes time $O(m^{e-\delta})=O(m^{e-\delta/2})$.
\item Step 4 takes time
\[
O\big(N^{2k-\delta}\big)\ =\ O\Big(m^{(2+\frac{e}{2k})(2k-\delta)}\Big),
\]
and the exponent of $m$ is
\[
\Big(2+\frac{e}{2k}\Big)\big(2k-\delta\big)\ =\ e+4k-2\delta-\frac{e\delta}{2k}\ \le\ e-\delta ,
\]
because $\frac{e\delta}{2k}\ge\frac{9k}{2}\ge 4k$ by the choice of $e$. So Step 4 takes time $O(m^{e-\delta})=O(m^{e-\delta/2})$.
\end{itemize}
Altogether, the algorithm decides $L$ in time $O(m^{e-\delta/2})$. This contradicts the choice of $L$ and proves \cref{fin:thm:main}.

\begin{remark}
Note that if $H$ were ``just'' $O(\log T)$ and not $\log T + O(1)$, then in the above argument we could have only concluded that $2^H = O(T^c) = O(m^{ec})$ for some constant $c \ge 1$. This would only prove that the resulting run-time of our algorithm for $L$ is $O(m^{ec-\delta/2})$. For $c > 1$, this might be more than $O(m^e)$, hence in that case we get no contradiction from the time hierarchy theorem.
\end{remark}

\begin{remark}
    We note that improving the exponent $2$ of our lower bound for pushdown reachability would be a major breakthrough: a lower bound of $N^{2+\epsilon}$ for some $\epsilon > 0$ on multitape Turing machines would imply that SAT cannot be solved in time $O(n^{1+\epsilon/3})$ on multitape Turing machines. Indeed, pushdown reachability can be solved in \emph{non-deterministic} time $N^2$~\cite{CMS22} and it is known that we can reduce non-deterministic computations of $T$ steps to SAT instances of size $T\cdot(\log T)^{O(1)}$~\cite{Cook88}. Hence, an $O(n^{1+\epsilon/3})$ algorithm for SAT would solve pushdown reachability in time $O(N^{2+\epsilon})$ and therefore, a lower bound of $O(N^{2+\epsilon})$ will imply an $O(n^{1+\epsilon/3})$ lower bound for SAT.
\end{remark}

\section{Conclusion}
\label{sec:conclusion}

We show the first unconditional super-linear lower bound for pushdown reachability on multitape Turing machines and hence, by reduction, for many other problems in program analysis. Our techniques use various ingredients, such as compressing Turing machine runs into locally addressable circuits with a short address length, storing only parts of this address in the state of a PDA, and encoding these parts by numbers that contain no three-term arithmetic progression so that they can be ``rearranged'' without much extra memory. Somewhat pleasantly, our proof does not use any new technique from theoretical computer science not already available since the 70s; however it does use the construction of 3-AP free sets, which is quite novel in connection to pushdown automata. Moreover, as we have pointed out, our proof works because of caring about the particular constants that appear in the $O(\cdot)$ notation. This suggests that, to get unconditional lower bounds for other problems, we have to start concentrating on specific details of reductions, such as the precise constants that we produce.

\subsection*{AI Methodology}

We acknowledge the use of various Generative AI tools (including Anthropic Fable 5.1 and 5.2, GPT 6.1, 6.2 Sol and 6.2 Astra) in the preparation of this paper. In particular, GPT 6.1-Sol found the proof of a weaker lower bound of $N^{3/2}$ completely by itself, after asking for an unconditional lower bound for the intersection non-emptiness problem. With assistance from Fable 5.1 and 5.2, the authors then understood, structured and polished this proof. During this period, GPT primarily seemed to be stuck on the $N^{3/2}$ lower bound. When the authors understood that proof, they were able to ask better questions to GPT to redirect the search for a $N^2$ lower bound. Once that was done, GPT 6.2 Astra found the technique to construct 3-AP free sets which led to the current $N^2$ lower bound. A (subjective) observation, similar to the one observed by others (for example in~\cite{WilliamsZ26}) from this project was the following: Sometimes current models are not able to make progress, until the existing ideas are understood and presented to them in a simplified form, after which they suddenly make great progress again.

Regarding writing and exposition, the content in this paper was revised by the authors after the initial draft of the proof by GPT-Sol and GPT-Astra, with assistance from Fable. We also used both these tools for minor rewrites, fixing typographical and grammatical mistakes and for making the figures.

\newpage

\bibliographystyle{plainurl}
\bibliography{references}

@article{Cook88,
  author  = {Cook, Stephen A.},
  title   = {Short Propositional Formulas Represent Nondeterministic Computations},
  journal = {Information Processing Letters},
  volume  = {26},
  number  = {5},
  pages   = {269--270},
  year    = {1988},
  doi     = {10.1016/0020-0190(88)90152-4}
}

@article{Behrend46,
	author  = {Behrend, F. A.},
	title   = {On sets of integers which contain no three terms in arithmetical progression},
	journal = {Proceedings of the National Academy of Sciences of the United States of America},
	volume  = {32},
	number  = {12},
	pages   = {331--332},
	year    = {1946},
	doi     = {10.1073/pnas.32.12.331}
}

@misc{ETHOpenAI,
  author = "OpenAI",
  title = "The exponential time hypothesis is false",
  url = "https://github.com/openai/math/tree/main/preprints/The-Exponential-Time-Hypothesis-is-False-October-25-2026",
  year = 2026
}

@misc{ETHAnthropic,
  author = "Anthropic",
  title = "A $@^{o(n)}$ time algorithm for 3-{SAT}",
  url = "https://github.com/anthropics/formal-math/blob/main/eth/paper.pdf",
  year = 2026
}

@article{Cook71,
  author       = {Stephen A. Cook},
  title        = {Characterizations of Pushdown Machines in Terms of Time-Bounded Computers},
  journal      = {J. {ACM}},
  volume       = {18},
  number       = {1},
  pages        = {4--18},
  year         = {1971},
  url          = {https://doi.org/10.1145/321623.321625},
  doi          = {10.1145/321623.321625},
  bibsource    = {dblp computer science bibliography, https://dblp.org}
}

@article{Grandjean88,
  author       = {Etienne Grandjean},
  title        = {A Natural NP-Complete Problem with a Nontrivial Lower Nound},
  journal      = {{SIAM} J. Comput.},
  volume       = {17},
  number       = {4},
  pages        = {786--809},
  year         = {1988},
  url          = {https://doi.org/10.1137/0217050},
  doi          = {10.1137/0217050},
  bibsource    = {dblp computer science bibliography, https://dblp.org}
}

@article{Grandjean90,
  author       = {Etienne Grandjean},
  title        = {A Nontrivial Lower Bound for an {NP} Problem on Automata},
  journal      = {{SIAM} J. Comput.},
  volume       = {19},
  number       = {3},
  pages        = {438--451},
  year         = {1990},
  url          = {https://doi.org/10.1137/0219028},
  doi          = {10.1137/0219028},
  bibsource    = {dblp computer science bibliography, https://dblp.org}
}

@inproceedings{Regan95,
  author       = {Kenneth W. Regan},
  title        = {On Superlinear Lower Bounds in Complexity Theory},
  booktitle    = {Proceedings of the Tenth Annual Structure in Complexity Theory Conference,
                  Minneapolis, Minnesota, USA, June 19-22, 1995},
  pages        = {50--64},
  publisher    = {{IEEE} Computer Society},
  year         = {1995},
  url          = {https://doi.org/10.1109/SCT.1995.514727},
  doi          = {10.1109/SCT.1995.514727},
  bibsource    = {dblp computer science bibliography, https://dblp.org}
}

@article{HartmanisStearns65,
  author = {Juris Hartmanis and Richard E. Stearns},
  title = {On the Computational Complexity of Algorithms},
  journal = {Transactions of the American Mathematical Society},
  volume = {117},
  pages = {285--306},
  year = {1965},
  doi = {10.1090/S0002-9947-1965-0170805-7}
}

@inproceedings{PPST83,
  author = {Wolfgang J. Paul and Nicholas Pippenger and Endre Szemer{\'e}di
            and William T. Trotter},
  title = {On Determinism Versus Non-Determinism and Related Problems},
  booktitle = {24th Annual Symposium on Foundations of Computer Science (FOCS)},
  pages = {429--438},
  publisher = {IEEE Computer Society},
  year = {1983},
  url = {https://scholarship.claremont.edu/hmc_fac_pub/1043/}
}

@article{AIK84,
  author = {Akeo Adachi and Shigeki Iwata and Takumi Kasai},
  title = {Some Combinatorial Game Problems Require {$\Omega(n^k)$} Time},
  journal = {Journal of the ACM},
  volume = {31},
  number = {2},
  pages = {361--376},
  year = {1984},
  doi = {10.1145/62.322433}
}

@inproceedings{SW15,
  author = {Joseph Swernofsky and Michael Wehar},
  title = {On the Complexity of Intersecting Regular, Context-Free, and Tree Languages},
  booktitle = {42nd International Colloquium on Automata, Languages, and Programming (ICALP), Part II},
  series = {Lecture Notes in Computer Science},
  volume = {9135},
  pages = {414--426},
  publisher = {Springer},
  year = {2015},
  doi = {10.1007/978-3-662-47666-6_33},
  url = {https://michaelwehar.com/documents/LBL_Equivalence.pdf}
}

@misc{AW26,
  author = {Josh Alman and Virginia Vassilevska Williams},
  title = {Truly Subquadratic {3SUM} and Truly Subcubic {APSP} via Triangles
           in Sparse Lopsided Graphs},
  year = {2026},
  note = {arXiv:2610.06783},
  doi = {10.48550/arXiv.2610.06783},
  url = {https://arxiv.org/abs/2610.06783}
}

@inproceedings{RHS95,
  author = {Thomas Reps and Susan Horwitz and Mooly Sagiv},
  title = {Precise Interprocedural Dataflow Analysis via Graph Reachability},
  booktitle = {22nd ACM SIGPLAN-SIGACT Symposium on Principles of Programming Languages (POPL)},
  pages = {49--61},
  publisher = {ACM},
  year = {1995},
  doi = {10.1145/199448.199462},
  url = {https://research.cs.wisc.edu/wpis/papers/popl95.pdf}
}

@inproceedings{BEM97,
  author = {Ahmed Bouajjani and Javier Esparza and Oded Maler},
  title = {Reachability Analysis of Pushdown Automata: Application to Model-Checking},
  booktitle = {8th International Conference on Concurrency Theory (CONCUR)},
  series = {Lecture Notes in Computer Science},
  volume = {1243},
  pages = {135--150},
  publisher = {Springer},
  year = {1997},
  url = {https://archive.model.in.tum.de/um/bibdb/info/esparza.BEM97.shtml.html}
}

@misc{Majumdar26,
      title={Beating {R}ytter's Bound: All-Pairs {CFL} Reachability in ${O}(n^3(\log\log n)^2/(\log n)^2)$ Time}, 
      author={Rupak Majumdar},
      year={2026},
      eprint={2611.12315},
      archivePrefix={arXiv},
      primaryClass={cs.PL},
      url={https://arxiv.org/abs/2611.12315}, 
}

@misc{WilliamsZ26,
      title={The k-Clique hypothesis is dead, for good}, 
      author={Virginia Vassilevska Williams and Or Zamir},
      year={2026},
      eprint={2611.09764},
      archivePrefix={arXiv},
      primaryClass={cs.DS},
      url={https://arxiv.org/abs/2611.09764}, 
}

@inproceedings{NFA,
  author    = {Karl Bringmann and Allan Gr{\o}nlund and Marvin K{\"u}nnemann
               and Kasper Green Larsen},
  title     = {The {NFA} Acceptance Hypothesis: Non-Combinatorial and Dynamic
               Lower Bounds},
  booktitle = {Proceedings of the Innovations in Theoretical Computer Science
               Conference (ITCS)},
  year      = {2024},
  note      = {Full version at \url{https://arxiv.org/abs/2311.10204}}
}

@article{Rytter85,
  author  = {Wojciech Rytter},
  title   = {Fast Recognition of Pushdown Automaton and Context-Free Languages},
  journal = {Information and Control},
  volume  = {67},
  number  = {1--3},
  pages   = {12--22},
  year    = {1985},
  doi     = {10.1016/S0019-9958(85)80024-3}
}

@inproceedings{Chaudhuri08,
  author = {Swarat Chaudhuri},
  title = {Subcubic Algorithms for Recursive State Machines},
  booktitle = {35th ACM SIGPLAN-SIGACT Symposium on Principles of Programming Languages (POPL)},
  pages = {159--169},
  publisher = {ACM},
  year = {2008},
  doi = {10.1145/1328438.1328460}
}

@article{CMS22,
  author = {Dmitry Chistikov and Rupak Majumdar and Philipp Schepper},
  title = {Subcubic Certificates for {CFL} Reachability},
  journal = {Proceedings of the ACM on Programming Languages},
  volume = {6},
  number = {POPL},
  pages = {41:1--41:29},
  year = {2022},
  doi = {10.1145/3498702},
  url = {https://arxiv.org/abs/2102.13095}
}

@inproceedings{BCM25,
  author = {A. R. Balasubramanian and Dmitry Chistikov and Rupak Majumdar},
  title = {Pushdown Model Checking above the Cubic Bottleneck},
  booktitle = {40th Annual ACM/IEEE Symposium on Logic in Computer Science (LICS)},
  pages = {622--635},
  publisher = {IEEE},
  year = {2025},
  doi = {10.1109/LICS65433.2025.00053},
  note = {Full version: arXiv:2608.24601, 2026},
  url = {https://arxiv.org/abs/2608.24601}
}

@inproceedings{Williams25,
  author = {R. Ryan Williams},
  title = {Simulating Time with Square-Root Space},
  booktitle = {57th Annual ACM Symposium on Theory of Computing (STOC)},
  publisher = {ACM},
  year = {2025},
  doi = {10.1145/3717823.3718225},
  url = {https://arxiv.org/abs/2502.17779}
}

@misc{Wehar26,
  author = {Michael Wehar},
  title = {Unconditional Time and Space Complexity Lower Bounds for Intersection Non-Emptiness},
  year = {2026},
  note = {arXiv:2512.00297, version 2; first posted in 2025},
  doi = {10.48550/arXiv.2512.00297},
  url = {https://arxiv.org/abs/2512.00297v2}
}

@article{PF79,
  author = {Nicholas Pippenger and Michael J. Fischer},
  title = {Relations among Complexity Measures},
  journal = {Journal of the ACM},
  volume = {26},
  number = {2},
  pages = {361--381},
  year = {1979},
  doi = {10.1145/322123.322138}
}

@techreport{Sha25,
  author = {Yakov Shalunov},
  title = {Improved Bounds on the Space Complexity of Circuit Evaluation},
  institution = {Electronic Colloquium on Computational Complexity},
  number = {TR25-078},
  year = {2025},
  url = {https://eccc.weizmann.ac.il/report/2025/078/}
}

@article{ABEGRY05,
  author = {Rajeev Alur and Michael Benedikt and Kousha Etessami and
            Patrice Godefroid and Thomas Reps and Mihalis Yannakakis},
  title = {Analysis of Recursive State Machines},
  journal = {ACM Transactions on Programming Languages and Systems},
  volume = {27},
  number = {4},
  pages = {786--818},
  year = {2005},
  doi = {10.1145/1075382.1075387},
  url = {https://www.cis.upenn.edu/~alur/toplas2005.pdf}
}

@inproceedings{HeintzeMcAllester,
  author    = {Nevin Heintze and
               David A. McAllester},
  title     = {On the Cubic Bottleneck in Subtyping and Flow Analysis},
  booktitle = {Proceedings, 12th Annual {IEEE} Symposium on Logic in Computer Science ({LICS}),
               Warsaw, Poland, June 29 - July 2, 1997},
  pages     = {342--351},
  publisher = {{IEEE} Computer Society},
  year      = {1997},
  url       = {https://doi.org/10.1109/LICS.1997.614960},
  doi       = {10.1109/LICS.1997.614960},
}

@unpublished{Neal,
  author = "Neal, Radford",
  year = "1989",
  title = "The computational complexity of taxonomic inference",
  note = "Unpublished manuscript. Available at \url{http://www.cs.toronto.edu/~radford/ftp/taxc.pdf}"
}

\appendix


\end{document}